# 3D atomic structure determination with ultrashort-pulse MeV electron diffraction

**Authors**

Vincent Hennicke[1]†, Max Hachmann[2]†, Paul Benjamin Klar[3]†, Patrick Y.A. Reinke[1], Tim Pakendorf[1], Jan Meyer[1], Hossein Delsim-Hashemi[2], Miriam Barthelmess[1], Sreevidya Thekku Veedu[1], Pontus Fischer[1], Ana C. Rodrigues[1], Arlinda Qelaj[2], Juna Wernsmann[2], Francois Lemery[2], Sebastian Günther[1], Sven Falke[1], Erik Fröjd[4], Aldo Mozzanica[4], Lukas Palatinus[5], Kai Rossnagel[6,7], Bernd Schmitt[4], Henry N. Chapman[1,8,9], Wim Leemans[2], Klaus Flöttmann[2]*, Alke Meents[1]*

**Affiliations**

[1]Center for Free-Electron Laser Science CFEL, Deutsches Elektronen-Synchrotron DESY; Notkestr. 85, 22607 Hamburg, Germany

[2]Deutsches Elektronen-Synchrotron DESY; Notkestr. 85, 22607 Hamburg, Germany

[3]Faculty of Geosciences and MAPEX Center for Materials and Processes, University of Bremen; Klagenfurter Str. 2, 28359 Bremen, Germany.

[4]Paul-Scherrer Institut; Forschungsstrasse 111, 5323 Villigen, Switzerland

[5]Institute of Physics of the Czech Academy of Sciences; Na Slovance 2, 182 21 Prague 8, Czechia.

[6]Institut für Experimentelle und Angewandte Physik, Christian-Albrechts-Universität zu Kiel; Olshausenstr. 40, 24098 Kiel, Germany.

[7]Ruprecht Haensel Laboratory, Deutsches Elektronen-Synchrotron DESY; Notkestr. 85, 22607 Hamburg, Germany.

[8]The Hamburg Centre for Ultrafast Imaging; Luruper Chaussee 149, 22761 Hamburg, Germany.

[9]Department of Physics, University of Hamburg; Luruper Chaussee 149, 22761 Hamburg, Germany.

* Corresponding authors. Email: klaus.floettmann@desy.de; alke.meents@desy.de

†These authors contributed equally to this work

**Abstract**

Understanding structure at the atomic scale is fundamental for the development of materials with improved properties. Compared to other probes providing atomic resolution, electrons offer the strongest interaction in combination with minimal radiation damage. Here, we report the successful implementation of MeV electron diffraction for ab initio 3D structure determination at atomic resolution. Using ultrashort electron pulses from the REGAE accelerator, we obtained high-quality diffraction data from muscovite and $1T$-TaS$_2$, enabling structure refinements according to the dynamical scattering theory and the accurate determination of hydrogen atom positions. The increased penetration depth of MeV electrons allows for structure determination from samples significantly thicker than those typically applicable in electron diffraction. These findings establish MeV electron diffraction as a viable approach for investigating a broad range of materials, including nanostructures and radiation-sensitive compounds, and open up new opportunities for *in-situ* and time-resolved experiments.

**MAIN TEXT**

Understanding the three-dimensional structure of a material is crucial for unraveling its physical properties. Diffraction experiments using X-rays, neutrons, or electrons are particularly well-suited for determining structures with (sub)-atomic resolution.

Among these methods, electrons stand out as a superior probe due to their significantly stronger elastic cross section and their ability to cause considerably less radiation damage to samples compared to X-rays (*1*). Unlike X-rays, which primarily interact with electrons of the sample, electrons interact with the electrostatic potential. This interaction enhances the visibility of light elements, particularly hydrogen atoms (*2*). Furthermore, highly brilliant electron beams are relatively easy to produce and can be precisely focused into submicron spot sizes using electromagnetic lenses. These features make electrons a uniquely economical and effective tool for structural investigations of submicron-sized samples, especially radiation-sensitive low-Z materials. As a result, electron diffraction has emerged as a rapidly growing research field (*3*). Today, three-dimensional electron diffraction (3D ED), also known as microcrystal electron diffraction (MicroED), is routinely used for structure determinations from nanocrystals. This technique is now being further advanced with the use of specialized devices designed specifically for electron crystallography (*4–7*).

The strong elastic interaction between electrons and matter however, also leads to challenges in structure determination (*8*). In particular, multiple scattering of electrons in the sample leads to non-kinematical diffracted intensities $I_{hkl}$ so that the simple relationship with the squared structure factor amplitudes, $I_{hkl} \propto |F_{hkl}|^2$, does not hold. According to the well-established dynamical theory of diffraction there is a non-linear dependence of $I_{hkl}$ on experimental parameters such as the electron wavelength, crystal orientation, crystal shape, and crystal thickness. Independent of that, electrons lose energy in the sample due to inelastic scattering events, which restricts the sample size for electron diffraction experiments. Depending on the density and chemical composition of the sample, the maximum tolerable thickness is typically limited to a few tens of nanometers in the case of inorganic (high-Z) and a few hundred nanometers for organic (low-Z) samples. These dimensions fall far below the resolution limit of optical microscopes, meaning that sample selection and preparation must be performed blindly, without the possibility of a straightforward inspection before the experiment. This poses a significant limitation to the broader applicability of electron diffraction (ED) for structure determination (*9, 10*).

Utilizing higher-energy electrons overcomes this limitation. Compared to 200 keV electrons, 3.48 MeV electrons have less than one-third of the elastic cross-section, allowing for a more than threefold increase in permissible sample thickness (*1*) while significantly reducing the impact of multiple scattering. As a result, organic and biological samples—previously limited to a maximum thickness of around 500 nm in conventional TEM devices—can now reach thicknesses of approximately 1.75 µm or even more in diffraction experiments using MeV electrons (*11*). This increase in tolerable sample thickness significantly broadens the applicability of ED to a much wider range of samples. Sample preparation steps, such as crystal growth, sample grinding, and mounting samples on support structures, can now be performed under an optical microscope, similar to X-ray experiments.

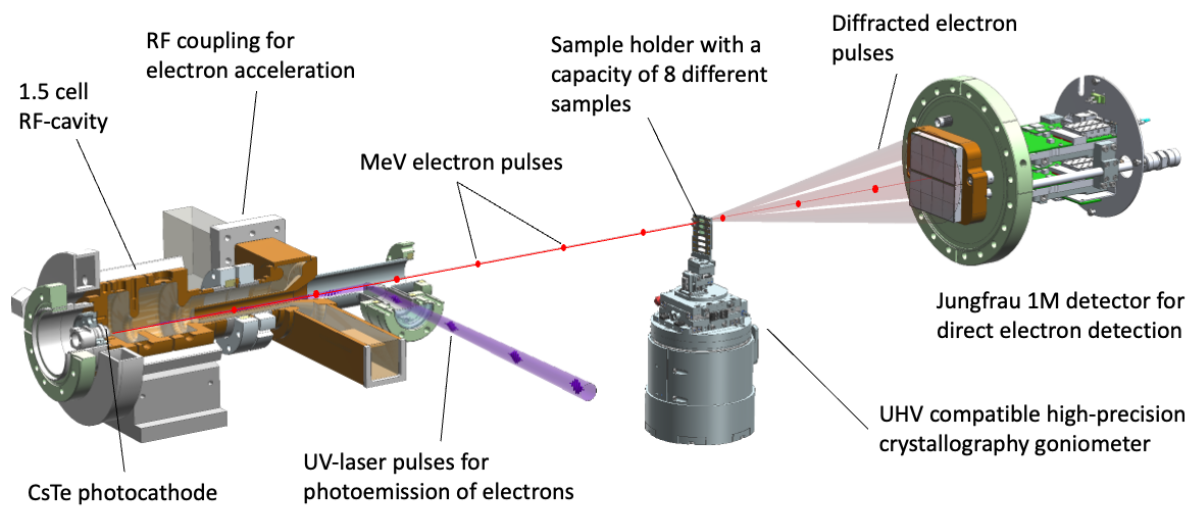

**Fig. 1. Concept of accelerator-based high energy electron diffraction with MeV electrons.** Electrons are emitted by illumination of the photocathode with short UV-pulses, which are then immediately accelerated to energies of 2 – 5 MeV in a 1.5 cell RF cavity with field gradients of up to 110 MV/m. The electron pulses (red dots) then interact with the sample, which is mounted on a single-axis goniometer for crystallographic data collection. Whilst the sample is rotated in the electron beam, diffraction patterns are recorded on a Jungfrau 1M integrating pixel detector.

Furthermore, the longer focal lengths in high-energy diffraction devices provide a larger working space around the sample enabling the use of a wide range of advanced and diverse sample environments. This setup allows for *in situ* experiments over a broad temperature range—from milli-Kelvin to several thousand Kelvin — as well as the use of different gas atmospheres, which is particularly advantageous for catalysis research. The sample holder and space for sample environments of a typical TEM, in contrast, is limited to dimensions in the millimeter range.

Moreover, high-energy MeV electrons are far less affected by electric and magnetic fields, enabling experiments that are difficult or impractical with lower-energy electrons; thereby further broadening their potential applications.

However, achieving electron energies of several MeV with the beam stability and coherence length necessary for high-quality structure determination in electron crystallography has been experimentally challenging. For instance, in a transmission electron microscope, accelerating electrons to 1 MeV in static DC fields requires a complex 36-stage accelerator tube (*12*).

## Accelerator-based high-energy electron diffraction

A more promising approach is the use of well-established radio frequency (RF)-based accelerator technology. Unlike conventional acceleration in a static field, RF-based accelerators deliver temporally short electron pulses with lengths varying between a few hundred picoseconds down to a few femtoseconds (*13, 14*). Such ultrashort probe pulses in combination with a short wavelength enable femtosecond time-resolved experiments with atomic resolution, which so far are predominantly conducted at much larger X-ray free-electron laser (XFEL) facilities (*15*).

Ultrashort, high-energy electron pulses have already proven highly effective in time-resolved ultrafast electron diffraction (UED) experiments. Beyond their application to solids and crystalline samples, UED techniques have also been successfully employed to study liquid and gaseous samples(*16–18*). However, the application of UED for 3D-structure determination at atomic resolution has been postulated to be unfeasible due to limitations in adequate sampling of reciprocal space (*19*). In fact, no diffraction experiments with ultrashort electron pulses have yet been reported that provided *ab initio* 3D structural information.

Here we report the first successful *ab initio* 3D structure determination at atomic resolution of the quasi-2D layered materials muscovite and the *1T* phase of tantalum disulfide with the relativistic electron gun for atomic exploration (REGAE) facility at DESY according to the concept of accelerator-based high energy electron diffraction (Fig 1). Muscovite is a layer silicate and has long been used as a very thin electrical insulator (*20, 21*). $1T$-$TaS_2$ belongs to the class of transition metal dichalcogenides (TMDCs) – a technologically highly relevant class of 2D quantum materials, which exhibit structurally very exciting properties such as charge density waves (CDWs) and undergo several phase transitions (*22–25*).

## REGAE accelerator facility

Diffraction experiments were performed at the 'Relativistic gun for atomic exploration – REGAE' facility at DESY in Hamburg, Germany (*26*). REGAE is a linear accelerator facility explicitly designed and built for time-resolved diffraction experiments with MeV electrons (Fig. S1). At REGAE, electrons are generated by photoemission inside an S-band (3 GHz) RF gun and rapidly accelerated to an energy of 3 to 5 MeV using field gradients of up to 110 MV/m (Fig. S2). REGAE is capable of generating electron pulses at a repetition rate of 50 Hz and with typical bunch charges of up to 100 fC with a transverse emittance as low as 10 nm (*27*).

For diffraction experiments REGAE is equipped with a dedicated UHV-compatible diffraction setup consisting of an in-line sample viewing microscope and a high-precision $e^-$-Roadrunner goniometer for crystallographic data collection (Fig. 1, S3, S4, S5, S6). The design was inspired by existing in-air X-ray diffraction setups commonly used at synchrotron sources (*28*).

## Direct electron detection

Diffraction data is recorded with an in-vacuum version of a Jungfrau 1M detector (Fig. 1, S7). The detector was originally developed for experiments with high-intensity X-ray pulses at XFELs (*29, 30*). In contrast to indirect, scintillator-based detectors, the Jungfrau detector directly records the electrical signal generated by the inelastic interactions of the electrons within the silicon sensor material, a signal that is proportional to the energy deposited in the sensor and which, given the high peak current, allowing a higher signal-to-noise ratio. In addition, a gating function records data only during a very short period centered around the arrival time of the electron pulse. All other signals, for example, dark current from the accelerator, arriving before and after the gating period, are not recorded and hence do not contribute to the background signal in the data.

The Jungfrau detector is composed of two 500 kpixel modules and can be operated at frame rates of up to 2 kHz. By using an automatic in-pixel gain selection with three different feedback capacitors, it provides a sufficient dynamic range for detecting strong Bragg reflections of up to 1200 electrons at 3.48 MeV per pixel per pulse.

**Muscovite and TaS$_2$ samples**

Muscovite, with chemical composition KAl$_3$Si$_3$O$_{10}$(OH)$_2$, crystallizes in space group *C*2/*c* with unit cell parameters $a = 5.21$ Å, $b = 9.04$ Å, $c = 20.03$ Å, and $\beta = 95.8°$ (*21*). Sheets of corner-sharing tetrahedra and edge-sharing octahedra are oriented perpendicular to crystallographic b-axis, resulting in extremely good cleavability between the layers. Muscovite samples were prepared by using the technique of exfoliation (Fig. S8).

The 1*T*-phase of TaS$_2$ used for our experiments is described in the space group *P*-3*m* with unit cell parameters of $a = 3.365$ Å and $c = 5.883$ Å (*22*). A charge density wave is formed by periodic, static displacements of Ta and S giving rise to additional Bragg satellite reflections in the diffraction patterns. The incommensurately modulated structure was previously described with the superspace group $X\bar{3}(\alpha, \beta, 0)0(-\alpha - \beta, \alpha, 0)0$ (*22*). 1*T*-TaS$_2$ samples were prepared by microtome cutting perpendicular to the naturally occurring layers oriented parallel to the crystallographic *ab*-plane (Fig. S9) (*23*).

**MeV electron diffraction data collection**

Electron diffraction data of muscovite and 1*T*-TaS$_2$ were recorded at REGAE at an energy of 3.48 MeV and a pulse duration of 600 fs at room temperature (293 K). For data collection, the samples were rotated from −60° to +60° in discrete steps of 0.01° for muscovite and from -65° to +65° in discrete steps of 0.005° for 1*T*-TaS$_2$, resulting in a pseudo-continuous rotation (Tab. S1). For each rotation increment, diffraction still images from 12 electron pulses were recorded. The diffraction images clearly show sharp Bragg reflections and very low noise levels between the reflections, highlighting the excellent achievable signal-to-noise ratio (Fig. 2). For 1*T*-TaS$_2$, weaker reflections around the Bragg peaks originating from the CDW can be clearly observed. The well-resolved spatial separation of the reflections is a clear indication of the high transverse and longitudinal coherence lengths of the electron bunches from REGAE.

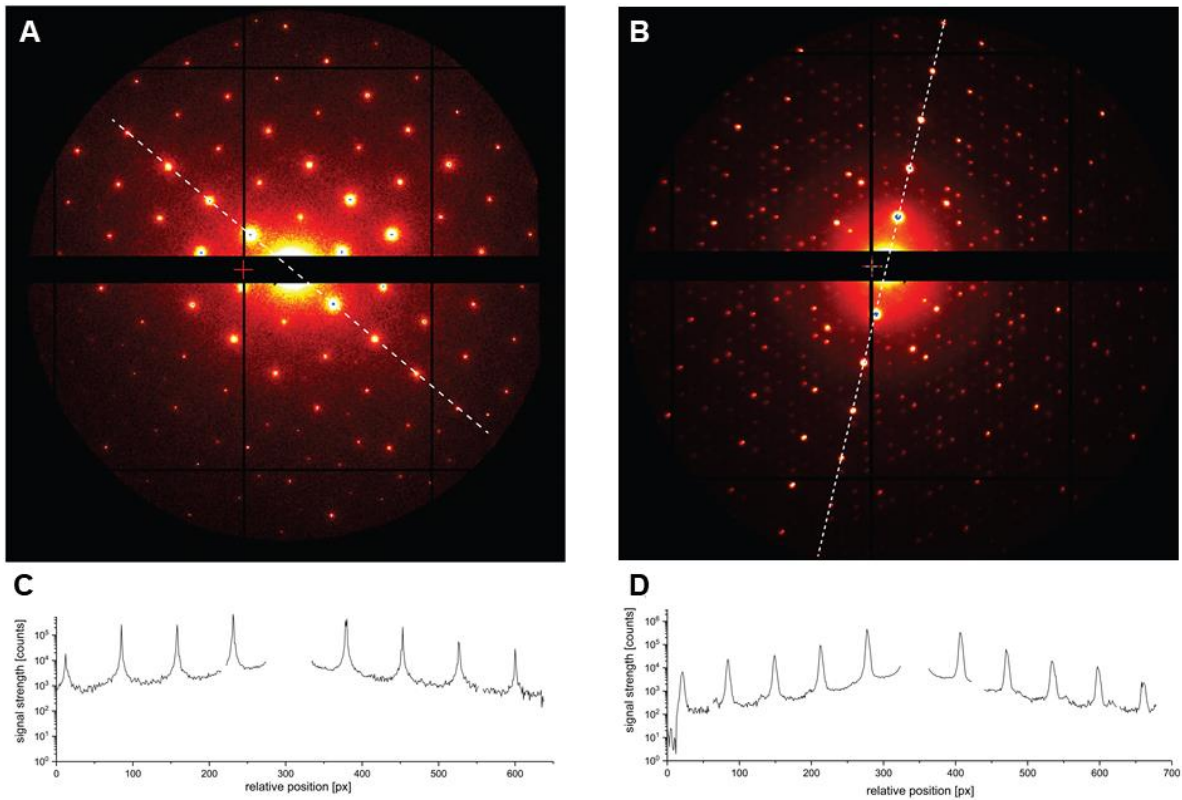

**Fig. 2. MeV electron diffraction patterns.** Sum of 12 single-shot images recorded with 3.48 MeV high-energy electrons at the REGAE facility from muscovite (**A**) and 1$T$-TaS$_2$ (**B**). The corresponding intensity profiles along the dashed lines indicated in the diffraction patterns highlight the achievable signal-to-noise-ratio resulting from the high coherence of the electron beam and the low background scattering signal (**C** and **D**).

With a bunch charge of 60 fC and a beam diameter of 500 µm, an integrated fluence of $2.8 \times 10^{-3}$ e$^-$ Å$^{-2}$ was delivered to the central part of the muscovite sample. The integrated fluence for the 1$T$-TaS$_2$ sample using pulse charges of less than 15 fC and a beam diameter of 50 µm was less than 0.14 e$^-$ Å$^{-2}$. During data collection, we could not detect any decay in the diffraction signals and quality, indicating an absence of radiation damage effects for the radiation doses applied here (Fig. 3A).

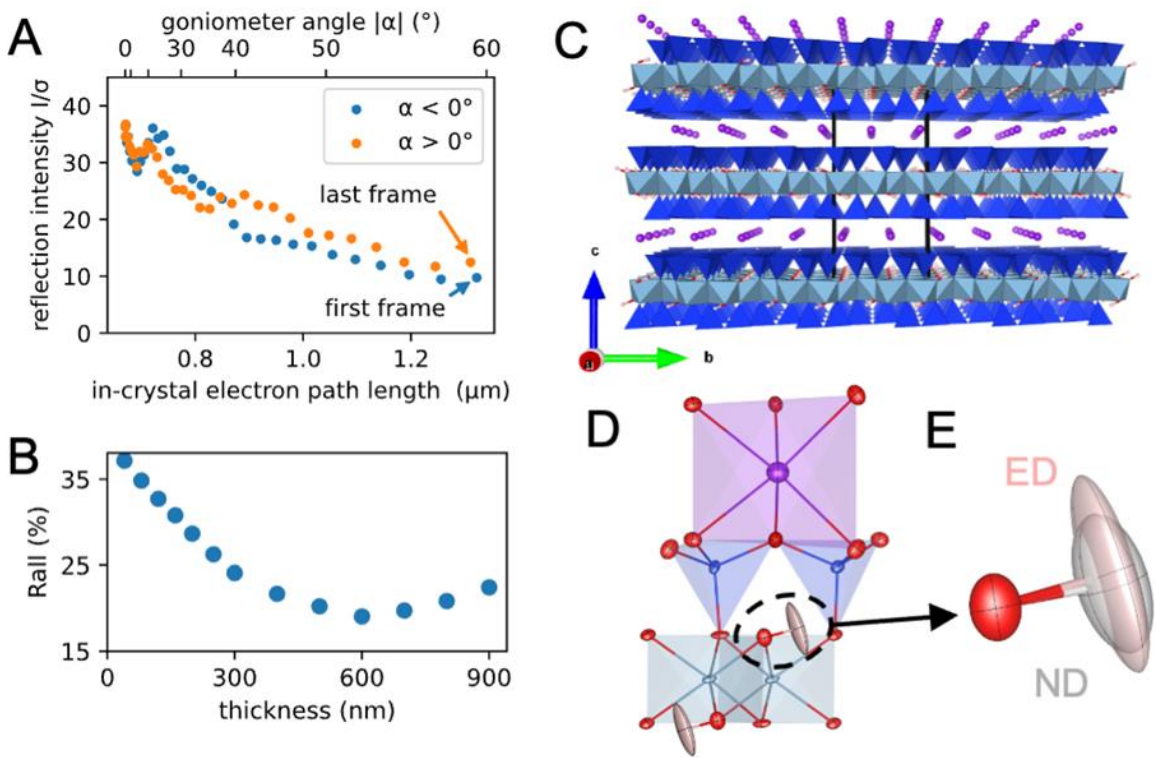

**Fig. 3**. **Results from dynamical structure refinement of the muscovite high-energy electron diffraction data.** (**A**) Decay of the mean I/σ(I) ratio in the diffraction data as function of the increasing effective sample thickness caused by sample rotation and the corresponding goniometer angle. (**B**) Dependence of the $R_{all}$ value from dynamical structure refinements as function of the adopted sample thickness. (**C**) View of the muscovite layer structure along the crystallographic a-axis. (**D**) Detailed view of the muscovite structure with anisotropic thermal displacement parameters indicated as ellipsoids (potassium = violet, aluminum = blue, oxygen = red, hydrogen = pink). (**E**) Overlay of the OH-group with anisotropic thermal displacement parameters shown as ellipsoids obtained with electron diffraction (ED, this work) and a reference model obtained with neutron diffraction (ND) (*21*).

**Structure solution and dynamical refinements**

Data reduction was performed with the software PETS2 (*31*). The quality of the diffraction datasets allowed a straightforward structure solution of the muscovite structure and the average 1*T*-TaS$_2$ structure using well-established software(*32–34*). The (3+2)d super space structure of 1*T*-TaS$_2$ could be solved with Superflip (*34*). Initial kinematical refinements, neglecting multiple scattering events, were performed with Jana2020 (*35*). They yielded $R_{all}$ values of 17.3% for muscovite and 25.5% for the modulated room temperature structure of 1*T*-TaS$_2$. To obtain a better agreement between the measured and calculated structure amplitudes and hence a better model of the structures, Jana2020 was also used to carry out dynamical structure refinements (*8, 36, 37*).

*Muscovite*

Dynamical refinement against the muscovite dataset resulted in an improved merged $R_{all}$ value of 9.4 % (Tab. S3). The refinement appeared highly sensitive to the assumed crystal thickness and crystal mosaicity. The $R_{all}$ values from the refinements show a distinct minimum for a thickness around 650 nm (Fig. 3). When freely refined, a crystal thickness of 634 nm is obtained (Fig 3A), which agrees well with the measured thickness of 670 nm. During data collection, the effective sample thickness, i.e. the length of the electron path through the crystal, increases up to 1.4 μm at a rotation angle of 60°. But even at this large thickness, an excellent data quality with a reasonable $I/\sigma(I) \cong 10$ could be obtained (Fig 3B).

In the resulting structural model, all details are well resolved and meaningful anisotropic thermal displacement parameters are obtained (Fig. 3C, 3D). A comparison of atom site coordinates with a reference structure of slightly different chemical composition (*21*) using the tool COMPSTRU yielded an average deviation of 0.032 Å with a maximum deviation of 0.083 Å for the non-hydrogen sites (*38*). The high quality of the diffraction data in combination with the high resolution in reciprocal space accessible with our method even allows the identification and free refinement of the hydrogen atom position with meaningful anisotropic thermal displacement parameters. The parameters obtained for the hydrogen atom and the resulting O-H bond length of 0.95(4) Å and the ADPs agree very well with neutron reference data, stating a length of 0.939(5) Å (Fig 3E, Tab. S4)(*21*). A comparative X-ray structure determination conducted from exactly the same sample as used for our MeV ED experiments yielded an O-H bond length of 0.81(6) Å and did not allow a free refinement of the ADPs within the independent atom model (Suppl. Text).

## *1T-TaS₂*

Dynamical refinement of the 1*T*-TaS$_2$ dataset resulted in an improved merged $R_{all}$ value of 4.3 % for the main reflections and 12.0 % for all reflections including the satellite reflections of 1*T*-TaS$_2$ (Tab. S3). Similar to muscovite, all structural details are well resolved, and meaningful thermal displacement parameters are obtained (Fig. 4A and B). The weak satellite reflections originating from the charge density wave were about 1-2 orders of magnitude lower in counts than the main structural reflections. Including these into the dynamical structure refinement enabled us to refine the incommensurate structure of 1*T*-TaS$_2$ (Fig. 4C). In agreement with a previous study conducted with X-rays from a 300-times thicker single crystal, an arrangement of star-like tantalum clusters within the layers is observed, each consisting of 13 tantalum atoms, (Fig. 4D, 4E) (*22*).

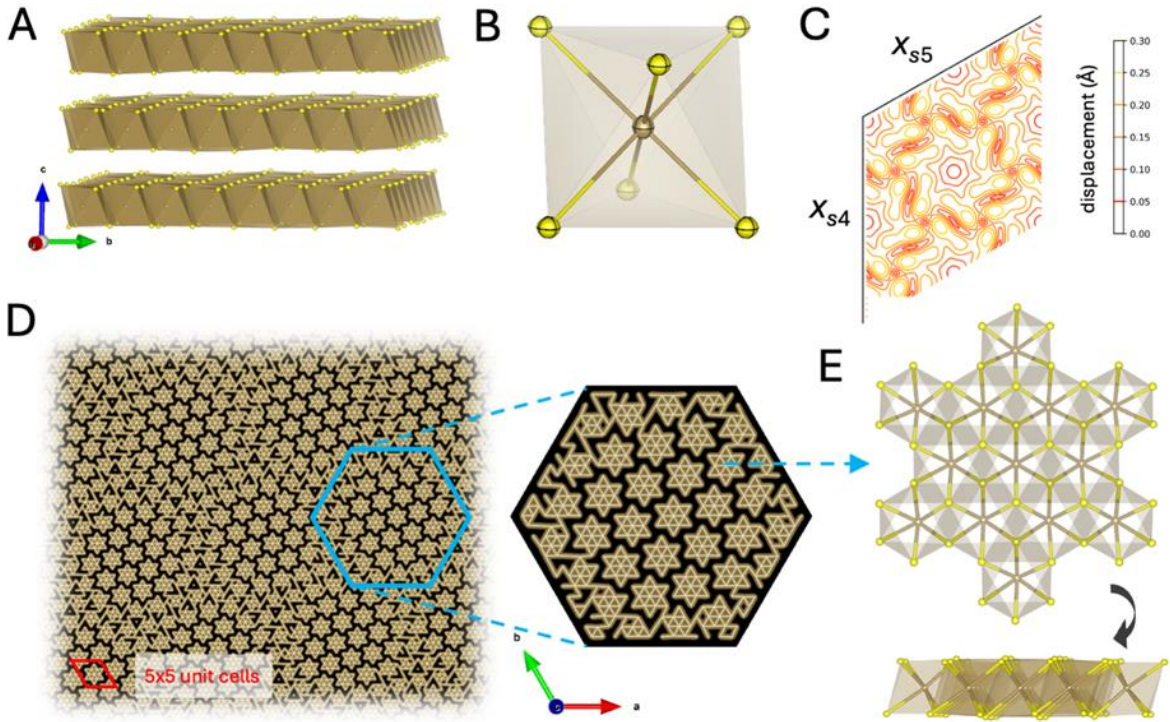

**Fig. 4. TaS₂ structure as determined by MeV electron diffraction at REGAE.** (**A**) TaS$_2$ layer structure perpendicular viewed along the crystallographic a-axis. (**B**) Octahedral coordination of the Ta-atom by 6 sulfur atoms with thermal displacement parameters represented as ellipsoids. (**C**) Refined modulation displacement amplitudes of Ta-atom as a function of $x_{s4}$ and $x_{s5}$ analogous to analysis by Spijkerman et al. (*22*). (**D**) Real space representation of the modulated structure of ~75 × 75 crystallographic unit cells showing the arrangement of star-shaped Ta-clusters, each consisting of 13 tantalum atoms. Ta atoms separated by less than 3.4 Å are connected by lines. Sulphur atoms are omitted. (**E**) View of a single star-shaped 13-atom tantalum cluster together with the coordinating sulphur atoms viewed along **c** (top) and rotated by 90° about **a** (bottom).

## Conclusions and Outlook

We have successfully applied accelerator-based high-energy electron diffraction to the field of 3D crystallographic structure determination, which has traditionally been dominated by X-rays. By integrating cutting-edge accelerator science with novel diffraction instrumentation, a direct electron-detection capable detector, optimization of dark current, and reduction of background scattering—paired with the latest software advancements for dynamical diffraction data analysis—we were able to solve and refine the structure of the quasi-2D layered materials muscovite and 1*T*-TaS$_2$ *ab initio* with dynamical refinements at very high quality as clearly indicated by very low merged R$_{all}$-values of 9.4 % for muscovite and 4.3 % for the unmodulated structure of TaS$_2$.

Electron energies exceeding 1 MeV—specifically 3.48 MeV in our case—strike an optimal balance between penetration depth and scattering contrast, making them ideal for structural studies of micro- and nanocrystals. For muscovite, the reduced elastic cross sections of high-energy electrons enabled us to acquire high-quality diffraction data from a 670 nm-thick sample, an achievement unattainable with current lab-based electron diffraction devices or electron microscopes. For proteins, which have a density approximately 2.5 times lower than muscovite, theoretical predictions and microscopy experiments on biological specimens (*39*) suggest that it should be possible to analyze crystals up to 2 μm in size. These dimensions are easily visible under an optical microscope, significantly simplifying and streamlining sample preparation processes.

At the same time, the elastic cross section at 3.48 MeV remains sufficiently high to enable high-quality structure determination from samples as thin as 30 nm. This capability allowed us to precisely resolve the incommensurate structure of 1$T$-TaS$_2$ using satellite reflections that are, on average, 1–2 orders of magnitude weaker than the main reflections. Given that the TaS$_2$ measurements utilized only a small fraction of the available bunch charge, it should be entirely feasible to perform diffraction experiments on just a few or even single layers of quantum materials at REGAE.

A current limitation of the method is the transversely large electron beam size caused by space charge effects, necessitating the use of laterally large yet relatively thin, pancake-like samples. To address this, we are implementing a bunch train mode at REGAE (*40*), in which the bunch charge of up to 100 fC is evenly distributed across 4,500 microbunches with a total duration of 1.5 µs. This approach enables the electron beam to be focused down to a few micrometers, allowing the investigation of isometric microcrystals of comparable size while preserving its excellent coherence properties.

For radiation-sensitive samples such as proteins, the larger penetration depth of MeV electrons, combined with their large scattering cross section and expected reduced radiation damage effects, should allow a routine collection of a complete dataset suitable for structure determination from a single micro- or nanocrystal, which is in most cases not possible with X-rays (*41*). This should make high-energy electrons the ideal probe for structure determination of such small crystals, for example in pharmaceutical compound screening experiments in the framework of structure-based drug discovery (*42*).

In addition, the more realistic hydrogen atom positions obtained in electron diffraction experiments (*2*), as shown here for muscovite, should lead to a better understanding of enzyme and other catalytic reactions, where hydrogen bonding and transfer reactions play a very important role (*43*).

One parameter not utilized in the present work is the high temporal resolution, down to the single-digit femtosecond range, achievable at linear accelerators such as REGAE. Currently, experiments that combine sub-atomic spatial resolution with femtosecond temporal resolution are predominantly conducted at XFEL sources. However, with significantly reduced radiation damage effects, such UED experiments at accelerator facilities could enable similar investigations without requiring continuous sample replenishment. This approach would, for instance, allow the study of coherent phonons and ultrafast phase transitions in single layers of 2D quantum materials like 1$T$-TaS$_2$ and other transition metal dichalcogenides with exceptional sensitivity (*44*).

With significantly lower investment and operating costs—particularly in terms of electricity consumption—accelerator-based electron diffraction experiments using MeV electrons present an economically and ecologically promising approach for structural and dynamical investigations of micro- and nanometer-sized materials. These methods serve as a valuable complement to existing microfocus synchrotron beamlines and XFELs.

# References

1. R. Henderson, The potential and limitations of neutrons, electrons and X-rays for atomic resolution microscopy of unstained biological molecules. *Q. Rev. Biophys.* **28**, 171–193 (1995).

2. L. Palatinus, P. Brázda, P. Boullay, O. Perez, M. Klementová, S. Petit, V. Eigner, M. Zaarour, S. Mintova, Hydrogen positions in single nanocrystals revealed by electron diffraction. *Science* **355**, 166–169 (2017).

3. T. Gruene, J. J. Holstein, G. H. Clever, B. Keppler, Establishing electron diffraction in chemical crystallography. *Nat. Rev. Chem.* **5**, 660–668 (2021).

4. B. L. Nannenga, D. Shi, A. G. W. Leslie, T. Gonen, High-resolution structure determination by continuous rotation data collection in MicroED. *Nat. Methods* **11**, 927–930 (2014).

5. M. Gemmi, E. Mugnaioli, T. E. Gorelik, U. Kolb, L. Palatinus, P. Boullay, S. Hovmöller, J. P. Abrahams, 3D Electron Diffraction: The Nanocrystallography Revolution. *ACS Cent. Sci.* **5**, 1315–1329 (2019).

6. S. Ito, F. J. White, E. Okunishi, Y. Aoyama, A. Yamano, H. Sato, J. D. Ferrara, M. Jasnowski, M. Meyer, Structure determination of small molecule compounds by an electron diffractometer for 3D ED/MicroED. *CrystEngComm* **23**, 8622–8630 (2021).

7. P. Simoncic, E. Romeijn, E. Hovestreydt, G. Steinfeld, G. Santiso-Quiñones, J. Merkelbach, Electron crystallography and dedicated electron-diffraction instrumentation. *Acta Crystallogr. Sect. E Crystallogr. Commun.* **79**, 410–422 (2023).

8. P. B. Klar, Y. Krysiak, H. Xu, G. Steciuk, J. Cho, X. Zou, L. Palatinus, Accurate structure models and absolute configuration determination using dynamical effects in continuous-rotation 3D electron diffraction data. *Nat. Chem.* **15**, 848–855 (2023).

9. H. M. E. Duyvesteyn, A. Kotecha, H. M. Ginn, C. W. Hecksel, E. V. Beale, F. de Haas, G. Evans, P. Zhang, W. Chiu, D. I. Stuart, Machining protein microcrystals for structure determination by electron diffraction. *Proc. Natl. Acad. Sci.* **115**, 9569–9573 (2018).

10. R. Bücker, P. Hogan-Lamarre, P. Mehrabi, E. C. Schulz, L. A. Bultema, Y. Gevorkov, W. Brehm, O. Yefanov, D. Oberthür, G. H. Kassier, R. J. Dwayne Miller, Serial protein crystallography in an electron microscope. *Nat. Commun.* **11**, 996 (2020).

11. J. J. Bozzola, L. D. Russell, *Electron Microscopy: Principles and Techniques for Biologists* (Jones and Bartlett, Boston, 2. ed. [Nachdr.], (2006). *The Jones and Bartlett series in biology*.

12. T. Kawasaki, I. Matsui, T. Yoshida, T. Katsuta, S. Hayashi, T. Onai, T. Furutsu, K. Myochin, M. Numata, H. Mogaki, M. Gorai, T. Akashi, O. Kamimura, T. Matsuda, N. Osakabe, A. Tonomura, K. Kitazawa, Development of a 1 MV field-emission transmission electron microscope. *J. Electron Microsc. (Tokyo)* **49**, 711–718 (2000).

13. S. P. Weathersby, G. Brown, M. Centurion, T. F. Chase, R. Coffee, J. Corbett, J. P. Eichner, J. C. Frisch, A. R. Fry, M. Gühr, N. Hartmann, C. Hast, R. Hettel, R. K. Jobe, E. N. Jongewaard, J. R. Lewandowski, R. K. Li, A. M. Lindenberg, I. Makasyuk, J. E. May, D. McCormick, M. N. Nguyen, A. H. Reid, X. Shen, K. Sokolowski-Tinten, T. Vecchione, S. L. Vetter, J. Wu, J. Yang, H. A. Dürr, X. J. Wang, Mega-electron-volt ultrafast electron diffraction at SLAC National Accelerator Laboratory. *Rev. Sci. Instrum.* **86**, 073702 (2015).

14. B. Zeitler, K. Floettmann, F. Grüner, Linearization of the longitudinal phase space without higher harmonic field. *Phys. Rev. Spec. Top. - Accel. Beams* **18**, 120102 (2015).

15. W. Decking, S. Abeghyan, P. Abramian, A. Abramsky, A. Aguirre, C. Albrecht, P. Alou, M. Altarelli, P. Altmann, K. Amyan, V. Anashin, E. Apostolov, K. Appel, D. Auguste, V. Ayvazyan, S. Baark, F. Babies, N. Baboi, P. Bak, V. Balandin, R. Baldinger, B. Baranasic, S. Barbanotti, O. Belikov, V. Belokurov, L. Belova, V. Belyakov, S. Berry, M. Bertucci, B. Beutner, A. Block, M. Blöcher, T. Böckmann, C. Bohm, M. Böhnert, V. Bondar, E. Bondarchuk, M. Bonezzi, P. Borowiec, C. Bösch, U. Bösenberg, A. Bosotti, R. Böspflug, M. Bousonville, E. Boyd, Y. Bozhko, A. Brand, J. Branlard, S. Briechle, F. Brinker, S. Brinker, R. Brinkmann, S. Brockhauser, O. Brovko, H. Brück, A. Brüdgam, L. Butkowski, T. Büttner, J. Calero, E. Castro-Carballo, G.

Cattalanotto, J. Charrier, J. Chen, A. Cherepenko, V. Cheskidov, M. Chiodini, A. Chong, S. Choroba, M. Chorowski, D. Churanov, W. Cichalewski, M. Clausen, W. Clement, C. Cloué, J. A. Cobos, N. Coppola, S. Cunis, K. Czuba, M. Czwalinna, B. D'Almagne, J. Dammann, H. Danared, A. de Zubiaurre Wagner, A. Delfs, T. Delfs, F. Dietrich, T. Dietrich, M. Dohlus, M. Dommach, A. Donat, X. Dong, N. Doynikov, M. Dressel, M. Duda, P. Duda, H. Eckoldt, W. Ehsan, J. Eidam, F. Eints, C. Engling, U. Englisch, A. Ermakov, K. Escherich, J. Eschke, E. Saldin, M. Faesing, A. Fallou, M. Felber, M. Fenner, B. Fernandes, J. M. Fernández, S. Feuker, K. Filippakopoulos, K. Floettmann, V. Fogel, M. Fontaine, A. Francés, I. F. Martin, W. Freund, T. Freyermuth, M. Friedland, L. Fröhlich, M. Fusetti, J. Fydrych, A. Gallas, O. García, L. Garcia-Tabares, G. Geloni, N. Gerasimova, C. Gerth, P. Geßler, V. Gharibyan, M. Gloor, J. Głowinkowski, A. Goessel, Z. Gołębiewski, N. Golubeva, W. Grabowski, W. Graeff, A. Grebentsov, M. Grecki, T. Grevsmuehl, M. Gross, U. Grosse-Wortmann, J. Grünert, S. Grunewald, P. Grzegory, G. Feng, H. Guler, G. Gusev, J. L. Gutierrez, L. Hagge, M. Hamberg, R. Hanneken, E. Harms, I. Hartl, A. Hauberg, S. Hauf, J. Hauschildt, J. Hauser, J. Havlicek, A. Hedqvist, N. Heidbrook, F. Hellberg, D. Henning, O. Hensler, T. Hermann, A. Hidvégi, M. Hierholzer, H. Hintz, F. Hoffmann, M. Hoffmann, M. Hoffmann, Y. Holler, M. Hüning, A. Ignatenko, M. Ilchen, A. Iluk, J. Iversen, J. Iversen, M. Izquierdo, L. Jachmann, N. Jardon, U. Jastrow, K. Jensch, J. Jensen, M. Jeżabek, M. Jidda, H. Jin, N. Johansson, R. Jonas, W. Kaabi, D. Kaefer, R. Kammering, H. Kapitza, S. Karabekyan, S. Karstensen, K. Kasprzak, V. Katalev, D. Keese, B. Keil, M. Kholopov, M. Killenberger, B. Kitaev, Y. Klimchenko, R. Klos, L. Knebel, A. Koch, M. Koepke, S. Köhler, W. Köhler, N. Kohlstrunk, Z. Konopkova, A. Konstantinov, W. Kook, W. Koprek, M. Körfer, O. Korth, A. Kosarev, K. Kosiński, D. Kostin, Y. Kot, A. Kotarba, T. Kozak, V. Kozak, R. Kramert, M. Krasilnikov, A. Krasnov, B. Krause, L. Kravchuk, O. Krebs, R. Kretschmer, J. Kreutzkamp, O. Kröplin, K. Krzysik, G. Kube, H. Kuehn, N. Kujala, V. Kulikov, V. Kuzminych, D. La Civita, M. Lacroix, T. Lamb, A. Lancetov, M. Larsson, D. Le Pinvidic, S. Lederer, T. Lensch, D. Lenz, A. Leuschner, F. Levenhagen, Y. Li, J. Liebing, L. Lilje, T. Limberg, D. Lipka, B. List, J. Liu, S. Liu, B. Lorbeer, J. Lorkiewicz, H. H. Lu, F. Ludwig, K. Machau, W. Maciocha, C. Madec, C. Magueur, C. Maiano, I. Maksimova, K. Malcher, T. Maltezopoulos, E. Mamoshkina, B. Manschwetus, F. Marcellini, G. Marinkovic, T. Martinez, H. Martirosyan, W. Maschmann, M. Maslov, A. Matheisen, U. Mavric, J. Meißner, K. Meissner, M. Messerschmidt, N. Meyners, G. Michalski, P. Michelato, N. Mildner, M. Moe, F. Moglia, C. Mohr, S. Mohr, W. Möller, M. Mommerz, L. Monaco, C. Montiel, M. Moretti, I. Morozov, P. Morozov, D. Mross, J. Mueller, C. Müller, J. Müller, K. Müller, J. Munilla, A. Münnich, V. Muratov, O. Napoly, B. Näser, N. Nefedov, R. Neumann, R. Neumann, N. Ngada, D. Noelle, F. Obier, I. Okunev, J. A. Oliver, M. Omet, A. Oppelt, A. Ottmar, M. Oublaid, C. Pagani, R. Paparella, V. Paramonov, C. Peitzmann, J. Penning, A. Perus, F. Peters, B. Petersen, A. Petrov, I. Petrov, S. Pfeiffer, J. Pflüger, S. Philipp, Y. Pienaud, P. Pierini, S. Pivovarov, M. Planas, E. Pławski, M. Pohl, J. Polinski, V. Popov, S. Prat, J. Prenting, G. Priebe, H. Pryschelski, K. Przygoda, E. Pyata, B. Racky, A. Rathjen, W. Ratuschni, S. Regnaud-Campderros, K. Rehlich, D. Reschke, C. Robson, J. Roever, M. Roggli, J. Rothenburg, E. Rusiński, R. Rybaniec, H. Sahling, M. Salmani, L. Samoylova, D. Sanzone, F. Saretzki, O. Sawlanski, J. Schaffran, H. Schlarb, M. Schlösser, V. Schlott, C. Schmidt, F. Schmidt-Foehre, M. Schmitz, M. Schmökel, T. Schnautz, E. Schneidmiller, M. Scholz, B. Schöneburg, J. Schultze, C. Schulz, A. Schwarz, J. Sekutowicz, D. Sellmann, E. Semenov, S. Serkez, D. Sertore, N. Shehzad, P. Shemarykin, L. Shi, M. Sienkiewicz, D. Sikora, M. Sikorski, A. Silenzi, C. Simon, W. Singer, X. Singer, H. Sinn, K. Sinram, N. Skvorodnev, P. Smirnow, T. Sommer, A. Sorokin, M. Stadler, M. Steckel, B. Steffen, N. Steinhau-Kühl, F. Stephan, M. Stodulski, M. Stolper, A. Sulimov, R. Susen, J. Świerblewski, C. Sydlo, E. Syresin, V. Sytchev, J. Szuba, N. Tesch, J. Thie, A. Thiebault, K. Tiedtke, D. Tischhauser, J. Tolkiehn, S. Tomin, F. Tonisch, F. Toral, I. Torbin, A. Trapp, D. Treyer, G. Trowitzsch, T. Trublet, T. Tschentscher, F. Ullrich, M. Vannoni, P. Varela, G. Varghese, G. Vashchenko, M. Vasic, C. Vazquez-Velez, A. Verguet, S. Vilcins-Czvitkovits, R. Villanueva, B. Visentin, M. Viti, E. Vogel, E. Volobuev, R. Wagner, N. Walker, T. Wamsat, H. Weddig, G. Weichert, H. Weise, R. Wenndorf, M. Werner, R. Wichmann, C. Wiebers, M. Wiencek, T. Wilksen, I. Will, L. Winkelmann, M. Winkowski, K. Wittenburg, A. Witzig, P. Wlk, T. Wohlenberg, M. Wojciechowski, F. Wolff-Fabris, G. Wrochna, K. Wrona, M. Yakopov, B. Yang, F. Yang, M. Yurkov, I. Zagorodnov, P. Zalden, A. Zavadtsev, D. Zavadtsev, A. Zhirnov, A. Zhukov, V. Ziemann, A. Zolotov, N. Zolotukhina, F. Zummack, D. Zybin, A MHz-repetition-rate hard X-ray free-electron laser driven by a superconducting linear accelerator. *Nat. Photonics* **14**, 391–397 (2020).

16. A. Kogar, A. Zong, P. E. Dolgirev, X. Shen, J. Straquadine, Y.-Q. Bie, X. Wang, T. Rohwer, I.-C. Tung, Y. Yang, R. Li, J. Yang, S. Weathersby, S. Park, M. E. Kozina, E. J. Sie, H. Wen, P. Jarillo-Herrero, I. R. Fisher, X. Wang, N. Gedik, Light-induced charge density wave in LaTe3. *Nat. Phys.* **16**, 159–163 (2020).

17. M.-F. Lin, N. Singh, S. Liang, M. Mo, J. P. F. Nunes, K. Ledbetter, J. Yang, M. Kozina, S. Weathersby, X. Shen, A. A. Cordones, T. J. A. Wolf, C. D. Pemmaraju, M. Ihme, X. J. Wang, Imaging the short-lived hydroxyl-hydronium pair in ionized liquid water. *Science* **374**, 92–95 (2021).

18. J. P. F. Nunes, K. Ledbetter, M. Lin, M. Kozina, D. P. DePonte, E. Biasin, M. Centurion, C. J. Crissman, M. Dunning, S. Guillet, K. Jobe, Y. Liu, M. Mo, X. Shen, R. Sublett, S. Weathersby, C. Yoneda, T. J. A. Wolf, J. Yang, A. A. Cordones, X. J. Wang, Liquid-phase mega-electron-volt ultrafast electron diffraction. *Struct. Dyn.* **7**, 024301 (2020).

19. T. Ishikawa, S. A. Hayes, S. Keskin, G. Corthey, M. Hada, K. Pichugin, A. Marx, J. Hirscht, K. Shionuma, K. Onda, Y. Okimoto, S. Koshihara, T. Yamamoto, H. Cui, M. Nomura, Y. Oshima, M. Abdel-Jawad, R. Kato, R. J. D. Miller, Direct observation of collective modes coupled to molecular orbital–driven charge transfer. *Science* **350**, 1501–1505 (2015).

20. J.-I. Liang, F. Hawthorne, I. Swainson, Triclinic muscovite: X-ray diffraction, neutron diffraction and photo-acoustic FTIR spectroscopy. *Can. Mineral.* **36**, 1017–1027 (1998).

21. G. D. Gatta, G. J. McIntyre, R. Sassi, N. Rotiroti, A. Pavese, Hydrogen-bond and cation partitioning in muscovite: A single-crystal neutron-diffraction study at 295 and 20 K. *Am. Mineral.* **96**, 34–41 (2011).

22. A. Spijkerman, J. L. de Boer, A. Meetsma, G. A. Wiegers, S. van Smaalen, X-ray crystal-structure refinement of the nearly commensurate phase of $1T-TaS2$ in (3+2)-dimensional superspace. *Phys. Rev. B* **56**, 13757–13767 (1997).

23. K. Haupt, M. Eichberger, N. Erasmus, A. Rohwer, J. Demsar, K. Rossnagel, H. Schwoerer, Ultrafast Metamorphosis of a Complex Charge-Density Wave. *Phys. Rev. Lett.* **116**, 016402 (2016).

24. S. Manzeli, D. Ovchinnikov, D. Pasquier, O. V. Yazyev, A. Kis, 2D transition metal dichalcogenides. *Nat. Rev. Mater.* **2**, 1–15 (2017).

25. T. Domröse, T. Danz, S. F. Schaible, K. Rossnagel, S. V. Yalunin, C. Ropers, Light-induced hexatic state in a layered quantum material. *Nat. Mater.* **22**, 1345–1351 (2023).

26. S. Manz, A. Casandruc, D. Zhang, Y. Zhong, R. A. Loch, A. Marx, T. Hasegawa, L. C. Liu, S. Bayesteh, H. Delsim-Hashemi, M. Hoffmann, M. Felber, M. Hachmann, F. Mayet, J. Hirscht, S. Keskin, M. Hada, S. W. Epp, K. Flöttmann, R. J. D. Miller, Mapping atomic motions with ultrabright electrons: towards fundamental limits in space-time resolution. *Faraday Discuss.* **177**, 467–491 (2015).

27. M. Hachmann, K. Flöttmann, Measurement of ultra low transverse emittance at REGAE. *Nucl. Instrum. Methods Phys. Res. Sect. Accel. Spectrometers Detect. Assoc. Equip.* **829**, 318–320 (2016).

28. A. Burkhardt, T. Pakendorf, B. Reime, J. Meyer, P. Fischer, N. Stübe, S. Panneerselvam, O. Lorbeer, K. Stachnik, M. Warmer, P. Rödig, D. Göries, A. Meents, Status of the crystallography beamlines at PETRA III. *Eur. Phys. J. Plus* **131**, 56 (2016).

29. A. Mozzanica, M. Andrä, R. Barten, A. Bergamaschi, S. Chiriotti, M. Brückner, R. Dinapoli, E. Fröjdh, D. Greiffenberg, F. Leonarski, C. Lopez-Cuenca, D. Mezza, S. Redford, C. Ruder, B. Schmitt, X. Shi, D. Thattil, G. Tinti, S. Vetter, J. Zhang, The JUNGFRAU Detector for Applications at Synchrotron Light Sources and XFELs. *Synchrotron Radiat. News* **31**, 16–20 (2018).

30. F. Leonarski, S. Redford, A. Mozzanica, C. Lopez-Cuenca, E. Panepucci, K. Nass, D. Ozerov, L. Vera, V. Olieric, D. Buntschu, R. Schneider, G. Tinti, E. Froejdh, K. Diederichs, O. Bunk, B. Schmitt, M. Wang, Fast and accurate data collection for macromolecular crystallography using the JUNGFRAU detector. *Nat. Methods* **15**, 799–804 (2018).

31. L. Palatinus, P. Brázda, M. Jelínek, J. Hrdá, G. Steciuk, M. Klementová, Specifics of the data processing of precession electron diffraction tomography data and their implementation in the program PETS2.0. *Acta Crystallogr. Sect. B Struct. Sci. Cryst. Eng. Mater.* **75**, 512–522 (2019).

32. G. M. Sheldrick, A short history of SHELX. *Acta Crystallogr. A* **64**, 112–122 (2008).

33. M. C. Burla, R. Caliandro, B. Carrozzini, G. L. Cascarano, C. Cuocci, C. Giacovazzo, M. Mallamo, A. Mazzone, G. Polidori, Crystal structure determination and refinement via SIR2014. *J. Appl. Crystallogr.* **48**, 306–309 (2015).

34. L. Palatinus, G. Chapuis, SUPERFLIP – a computer program for the solution of crystal structures by charge flipping in arbitrary dimensions. *J. Appl. Crystallogr.* **40**, 786–790 (2007).

35. V. Petříček, L. Palatinus, J. Plášil, M. Dušek, Jana2020 – a new version of the crystallographic computing system Jana. *Z. Für Krist. - Cryst. Mater.* **238**, 271–282 (2023).

36. L. Palatinus, V. Petříček, C. A. Corrêa, Structure refinement using precession electron diffraction tomography and dynamical diffraction: theory and implementation. *Acta Crystallogr. Sect. Found. Adv.* **71**, 235–244 (2015).

37. L. Palatinus, C. A. Corrêa, G. Steciuk, D. Jacob, P. Roussel, P. Boullay, M. Klementová, M. Gemmi, J. Kopeček, M. C. Domeneghetti, F. Cámara, V. Petříček, Structure refinement using precession electron diffraction tomography and dynamical diffraction: tests on experimental data. *Acta Crystallogr. Sect. B Struct. Sci. Cryst. Eng. Mater.* **71**, 740–751 (2015).

38. G. de la Flor, D. Orobengoa, E. Tasci, J. M. Perez-Mato, M. I. Aroyo, Comparison of structures applying the tools available at the Bilbao Crystallographic Server. *J. Appl. Crystallogr.* **49**, 653–664 (2016).

39. V. E. Cosslett, High-Voltage Electron Microscopy. *Q. Rev. Biophys.* **2**, 95–133 (1969).

40. C. Mahnke, U. Grosse-Wortmann, M. Hachmann, V. Hennicke, C. Li, J. Meyer, T. Pakendorf, K. Flöttmann, A. Meents, I. Hartl, First demonstration of ultrafast-electron diffraction using a 3-GHz electro-optical comb generator as a UV photoinjector laser. *EPJ Web Conf.* **307**, 02030 (2024).

41. J. M. Holton, A beginner's guide to radiation damage. *J. Synchrotron Radiat.* **16**, 133–142 (2009).

42. S. Günther, P. Y. A. Reinke, Y. Fernández-García, J. Lieske, T. J. Lane, H. M. Ginn, F. H. M. Koua, C. Ehrt, W. Ewert, D. Oberthuer, O. Yefanov, S. Meier, K. Lorenzen, B. Krichel, J.-D. Kopicki, L. Gelisio, W. Brehm, I. Dunkel, B. Seychell, H. Gieseler, B. Norton-Baker, B. Escudero-Pérez, M. Domaracky, S. Saouane, A. Tolstikova, T. A. White, A. Hänle, M. Groessler, H. Fleckenstein, F. Trost, M. Galchenkova, Y. Gevorkov, C. Li, S. Awel, A. Peck, M. Barthelmess, F. Schlünzen, P. Lourdu Xavier, N. Werner, H. Andaleeb, N. Ullah, S. Falke, V. Srinivasan, B. A. França, M. Schwinzer, H. Brognaro, C. Rogers, D. Melo, J. J. Zaitseva-Doyle, J. Knoska, G. E. Peña-Murillo, A. R. Mashhour, V. Hennicke, P. Fischer, J. Hakanpää, J. Meyer, P. Gribbon, B. Ellinger, M. Kuzikov, M. Wolf, A. R. Beccari, G. Bourenkov, D. von Stetten, G. Pompidor, I. Bento, S. Panneerselvam, I. Karpics, T. R. Schneider, M. M. Garcia-Alai, S. Niebling, C. Günther, C. Schmidt, R. Schubert, H. Han, J. Boger, D. C. F. Monteiro, L. Zhang, X. Sun, J. Pletzer-Zelgert, J. Wollenhaupt, C. G. Feiler, M. S. Weiss, E.-C. Schulz, P. Mehrabi, K. Karničar, A. Usenik, J. Loboda, H. Tidow, A. Chari, R. Hilgenfeld, C. Uetrecht, R. Cox, A. Zaliani, T. Beck, M. Rarey, S. Günther, D. Turk, W. Hinrichs, H. N. Chapman, A. R. Pearson, C. Betzel, A. Meents, X-ray screening identifies active site and allosteric inhibitors of SARS-CoV-2 main protease. *Science* **372**, 642–646 (2021).

43. D. Herschlag, M. M. Pinney, Hydrogen Bonds: Simple after All? *Biochemistry* **57**, 3338–3352 (2018).

44. D. Filippetto, P. Musumeci, R. K. Li, B. J. Siwick, M. R. Otto, M. Centurion, J. P. F. Nunes, Ultrafast electron diffraction: Visualizing dynamic states of matter. *Rev. Mod. Phys.* **94**, 045004 (2022).

45. E. Fröjdh, F. Baruffaldi, A. Bergamaschi, M. Carulla, R. Dinapoli, D. Greiffenberg, J. Heymes, V. Hinger, R. Ischebeck, S. Mathisen, J. McKenzie, D. Mezza, K. Moustakas, A. Mozzanica, B. Schmitt, J. Zhang, Detection of MeV electrons using a charge integrating hybrid pixel detector. *J. Instrum.* **17**, C12004 (2022).

46. K. S. Novoselov, A. H. C. Neto, Two-dimensional crystals-based heterostructures: materials with tailored properties. *Phys. Scr.* **2012**, 014006 (2012).

47. S. Redford, M. Andrä, R. Barten, A. Bergamaschi, M. Brückner, R. Dinapoli, E. Fröjdh, D. Greiffenberg, C. Lopez-Cuenca, D. Mezza, A. Mozzanica, M. Ramilli, M. Ruat, C. Ruder, B. Schmitt, X. Shi, D. Thattil, G. Tinti, S. Vetter, J. Zhang, First full dynamic range calibration of the JUNGFRAU photon detector. *J. Instrum.* **13**, C01027 (2018).

48. M. Catti, G. Ferraris, S. Hull, A. Pavese, Powder neutron diffraction study of 2M1 muscovite at room pressure and at 2 GPa. *Eur. J. Mineral.*, 171–178 (1994).

49. P. Brázda, M. Klementová, Y. Krysiak, L. Palatinus, Accurate lattice parameters from 3D electron diffraction data. I. Optical distortions. *IUCrJ* **9**, 735–755 (2022).

50. M. Khouchen, P. B. Klar, H. Chintakindi, A. Suresh, L. Palatinus, Optimal estimated standard uncertainties of reflection intensities for kinematical refinement from 3D electron diffraction data. *Acta Crystallogr. Sect. Found. Adv.* **79**, 427–439 (2023).

51. G. M. Sheldrick, Crystal structure refinement with SHELXL. *Acta Crystallogr. Sect. C Struct. Chem.* **71**, 3–8 (2015).

52. W. Kabsch, XDS. *Acta Crystallogr. D Biol. Crystallogr.* **66**, 125–132 (2010).

53. L. J. Bourhis, O. V. Dolomanov, R. J. Gildea, J. A. K. Howard, H. Puschmann, The anatomy of a comprehensive constrained, restrained refinement program for the modern computing environment - Olex2 dissected. *Acta Crystallogr. Sect. Found. Adv.* **71**, 59–75 (2015).

54. G. de la Flor, D. Orobengoa, E. Tasci, J. M. Perez-Mato, M. I. Aroyo, Comparison of structures applying the tools available at the Bilbao Crystallographic Server. *J. Appl. Crystallogr.* **49**, 653–664 (2016).

**Acknowledgments**

We thank Robin Schubert and Iñaki de Diego for assistance with sample preparation. We also thank Oleksandr Yefanov, Alexandra Tolstikova and Marina Galchenkova for support with data handling. This research was supported in part through the Maxwell computational resources operated at DESY. We acknowledge Deutsches Elektronen-Synchrotron (DESY; Hamburg, Germany), a member of the Helmholtz Association HGF, for the provision of experimental facilities. Parts of this research were carried out at PETRA III at beamline P11.

**Funding**

Czech Science Foundation grant 21-05926X (LP)

German Research Foundation, Cluster of Excellence "CUI: Advanced Imaging of Matter" - EXC 2056 - project ID 390715994 (HNC)

German Federal Ministry of Education and Research, project "MHz-SFX", ID 13K20CHA (AM, HNC)

Helmholtz Association Impulse and Networking funds InternLabs-0011 "HIR3X" and project "FISVIR" (AM)

**Author contributions**

VH – Conceptualization, Investigation Resources; MH – Conceptualization, Investigation; PBK – Conceptualization, Formal analysis, Methodology, Writing – original draft, Writing – review & editing; PYAR – Formal analysis, Investigation; TP – Resources; JM – Resources, Software; HDH – Investigation, Methodology; MB – Resources; STV – Investigation; PF – Resources, Software; ACR – Investigation, Software; AQ – Resources; JW – Investigation; FL – Writing – review & editing; SG – Project administration, Writing – original draft, Writing – review & editing; SF – Project administration, Writing – original draft, Writing – review & editing; EF – Investigation, Resources; AMo – Investigation, Resources; LP – Formal analysis, Writing – review & editing; KR – Conceptualization,

Resources, Writing – review & editing; BS – Conceptualization, Methodology, Resources; HNC – Funding acquisition, Writing – review & editing; WL – Funding acquisition, Writing – review & editing; KF – Conceptualization, Methodology, Project administration, Writing – original draft, Writing – review & editing; AM – Conceptualization, Funding acquisition, Methodology, Project administration, Writing – original draft, Writing – review & editing;

**Competing interests**

Authors declare that they have no competing interests.

**Data and materials availability**

Diffraction data for the MeV-ED datasets for muscovite and 1$T$-TaS$_2$ and the X-ray diffraction dataset for muscovite are available from the authors upon request. All other data are available in the main text or the supplementary materials.

# Supplementary Materials for

**3D atomic structure determination with ultrashort-pulse MeV electron diffraction**

- Materials and Methods
- Supplementary Text
- Figs. S1 to S12
- Tables S1 to S8
- References *(45-53)*

**Materials and Methods**

REGAE accelerator

REGAE provides short pulses of electrons over an energy range of 3-5 MeV with bunch charges of up to 100 fC and pulse durations down to 20 fs (rms). Electrons are emitted from a photocathode upon irradiation with a short UV laser pulse and directly accelerated in a 3 GHz RF-gun (S-band) with field gradients of up to 110 MV/m on the cathode. These electron pulses with a typical duration of several hundreds of femtoseconds can then be further longitudinally compressed using a buncher cavity leading to pulse duration down to 20 fs (rms). A technical overview drawing of the REGAE facility is provided in figure S1.

Several collimation systems in the electron beam path allow to additionally shape the electron beam and scrape off halo electrons and dark-current electrons. The lengths of the REGAE accelerator part as measured from the photocathode to the sample position is 5.5 m (Fig. S2). With an additional propagation length of 5.43 m between the sample position and the detector the overall length of the REGAE experiment adds up to 10.9 m.

*Laser system*

A Titan-sapphire laser (model: Legend, vendor: Coherent) system is utilized for both electron generation in the photocathode and optionally also for time-resolved experiments for sample excitation in the UV/VIS range. For synchronization to the RF-system the oscillator cavity of the laser is synchronized to the master frequency of REGAE by a direct conversion system with a temporal stability of 11 fs (rms). After passing through the amplifier the 800 nm laser light is split into two beams. A small fraction (< 10%) of the beam is frequency-tripled and then directly guided to the photocathode for electron generation (probe pulse). The larger fraction can be utilized for sample excitation (pump pulse). The path lengths of the two laser beams defines the arrival time at the corresponding target (photocathode and sample). Typically, a shorter beam path is required for sample excitation, so that the sample is excited before arrival of the probing electron pulse. Variation of the path length of the pump laser beam allows to adjust the time delay between the two beams.

*Electron gun*

REGAE is equipped with a load lock system allowing for fast exchange of the photocathode und ultra-high vacuum (UHV). For the present experiment Mo was used as photocathode material. The RF gun is a 1.5 cell normal conducting copper cavity operated with up to 5 MW RF power and a field gradient of up to 110 MV/m. The temperature of the gun is stabilized to < 10 mK to reduce the gun contribution to the overall timing jitter. The UV laser light for photoemission is coupled-in at a small angle with respect to the beam axis using an in vacuum mirror located about 0.5 m downstream of the photocathode resulting in spot size of 10 μm (rms). The chosen combination of the laser spot size, cathode material, and field gradient results in a beam emittance of 20 nm (rms).

*Buncher cavity*

The buncher is a 4-cell S-band cavity located about 1.2 m downstream of the photocathode RF-gun. The beam passes on the zero-crossing of the field through the cavity so that a correlated energy spread is impinged onto the bunch leading to a longitudinal bunch compression in the following drift section towards the sample. The field amplitude is tuned such that the bunch length reaches a minimum at the sample position.

REGAE diffraction experiment

*Large experimental chamber*

For diffraction experiments REGAE is equipped with a large experimental UHV chamber (Fig. S3). The chamber is internally equipped with 4 heavy-load horizontal translation stages allowing to carry and precisely position experimental equipment in the electron beam path.

*Inline sample viewing microscope*

For sample visualization and beamline alignment an inline sample viewing microscope can be inserted into the electron beam path allowing to permanently visualize the sample with optical light also during the diffraction experiment. Its functioning principle is illustrated in Fig. S4. The inline microscope allows to view the sample with visible light colinear to the electron beam. For this, the electron beam passes through a 1.0 mm drill hole along the optical axis of the microscope objective. Shortly after the light exits the microscope it is deflected upwards by a 90-degree deflecting prism, which is equipped with a drill hole. A tube lense then focuses the light on a CCD camera. Non-UHV compatible parts of the device such as the CCD camera are placed in a sealed containment at atmospheric pressure, allowing to operate the device in a UHV environment. The on-axis microscope provides a field of view of 1250 x 1050 µm with an optical resolution of about 1 µm.

The 1.0 mm drill hole in the objective is typically fitted with a tantalum capillary with an inner diameter of 0.5 mm serving as a collimator for the electron beam leading to significant reduction of the background signal on the detector. The entire device is mounted on a motorized 5-axes positioning system allowing for x-, y-, and z-positioning and angular alignment with pitch and yaw. The inline viewing microscope was purchased from suna precision (www.suna-precision.com).

*High-precision $e^-$-Roadrunner crystallography goniometer*

For diffraction experiments with solid samples the chamber is equipped with an $e^-$-Roadrunner goniometer, a compact and UHV compatible single-axis crystallographic goniometer with a high-precision vertical rotation axis (Fig. S5). The goniometer axis itself consists of a directly servomotor-driven and UHV compatible rotation axis based on ceramic bearings providing 360-degree rotation capability with an angular resolution of 0.001° in combination with a sphere of confusion smaller than 1 µm. A centering stage mounted on top of the rotation axis provides travel ranges of +/- 6mm in x-, y-, and z-direction with a resolution < 1 µm and allows precise positioning of the sample in the center of the rotation axis. This assures that the sample remains in the electron beam during rotation for data collection. The goniometer axis itself is mounted on a 3-axes motorized translation stage in an orthogonal configuration allowing to position the rotation axis in x-, y-, and z-direction precisely at the position of the electron beam. The goniometer can carry different sample holders ranging from custom made rectangularly shaped silicon nitride membranes on a silicon frame to conventional round 3 mm diameter standard EM grids. The assembly of the on-axis microscope together with the $e^-$ Roadrunner goniometer and a sample holder is shown in figure S6.

*Jungfrau 1M pixel detector*

Diffraction images are recorded on a UHV compatible version of the Jungfrau 1M detector, which is placed 5.34 m downstream of the sample and terminating the vacuum system (Fig. S7). The detector has been originally developed for experiments with high-intensity X-ray pulses at X-ray free-electron lasers but can also be used for experiments at synchrotron radiation sources (*29*). The Jungfrau detector directly records the electrical signal generated by the inelastic interactions of the electrons with the silicon sensor material. This signal, which is proportional to the energy deposited in the sensor and given its strong intensity, allows achieving a higher signal over noise ratio than possible with commonly used indirect scintillator-based detection. Each 3.48 MeV electron impinging on the detector surface will behave as a minimum ionizing particle, ionizing e-h+ pairs along its mostly straight track and then exiting the back side of the sensor with most of its energy and momentum. It is then going to release the rest of the energy and stop in the downstream readout, which can withstand radiation doses up to a ~1 MGy, so that radiation damage for operating condition at REGAE is not expected. The Jungfrau 1M detector is composed of two 500 kpixel modules. The pixel size is 75 µm squared and the detector can be operated at frames rates of up to 2 kHz. The system is water-cooled at room temperature, and is operated at vacuum levels $< 3 \times 10^{-7}$ mbar for the measurements.

By using automatic in-pixel gain switching with three different feedback capacitors Jungfrau provides a sufficient dynamic range for detecting strong Bragg reflections while at the same time maintaining single electron resolution with a high signal to noise ratio. In our case with a 320 µm thick silicon sensor the dynamic range is around 1200 incident electrons per pixel and frame (Fröjdh 2022)(*45*).

Sample Preparation

*Muscovite*

The muscovite sample (mica, "rectangular #56 - 75 × 25 mm$^2$", quality "V1") was purchased from Plano GmbH (Germany). To obtain a clean surface the muscovite plate was exfoliated several times with adhesive tape until a flat surface was achieved (*46*) (Fig. S8). Subsequently, thin multilayer sheets with optically smooth and homogeneous surfaces and without obvious cracks were produced from the muscovite supply plate by exfoliation with adhesive tape again. These multilayer sheets were then glued over the apertures with a size $0.75 \times 3.00$ mm$^2$ of the REGAE silicon sample holders with dimensions of $5 \times 5$ mm$^2$ and thickness of 0.2 mm using ethyl-2-cyanoacrylate adhesive (UHU "mini blitzschnell" gel, UHU GmbH & Co KG; Fig. S6). After curing of the glue, the sample was further thinned down by additional exfoliation so that ideally a very thin sample consisting of few layers only was obtained. The sample used for the diffraction experiment spanned the entire aperture of the silicon sample holder with dimensions of $0.75 \times 3.0$ mm$^2$ and had a thickness of 670 nm as determined independently with scanning electron microscopy measurements at an edge of the sample. The large aperture size of the sample holder allows for large rotation angles without obstructing the electron beam by the support frame.

*Tantalum disulfide (1T-TaS$_2$)*

1$T$-TaS$_2$ bulk single crystals were grown from highest purity substrate elements by chemical vapor transport using iodine as transport agent (*46*). A plate-like single crystal showing well defined crystal edges and a smooth surface showing no major defects with dimensions of about $2 \times 2 \times 0.2$ mm$^3$ was selected. 1$T$-TaS$_2$ crystals exhibit good cleavability into layers perpendicular to the direction of the shortest crystal dimension, the crystallographic c-axis. Very thin sheets of 1$T$-TaS$_2$ with thicknesses down to 25 nm as required for experiments at REGAE were obtained by microtome cutting. For this, the crystal was first glued with epoxy glue onto the top of an epoxy resin block for subsequent microtome cutting (Fig. S9). The sample was then trimmed with a knife to a freestanding and well accessible rectangular shape. The block with the trimmed sample on top was then mounted on the microtome head. Cutting was performed using a LEICA, Ultramicrotome EM UC7 with feed steps of 30 nm and a knife speed of 6 mm s$^{-1}$. Subsequent to slicing, the sample swimming on the water surface was collected with SiN-coated silicon chips and was dried in the air. The sample flake used for the diffraction experiment had lateral dimensions of $0.9 \times 0.4$ mm$^2$ and a thickness of 30 nm. For the measurements the flake was mounted on a silicon frame equipped with rectangular apertures covered with 30 nm Si$_3$N$_4$ membranes.

Data collection

Crystallographic rotation data from the muscovite and 1$T$-TaS$_2$ samples were collected at REGAE facility at room temperature using an electron energy of 3.48 MeV. Before data collection both samples were centered in the goniometer rotation axis and in the electron using the inline-sample viewing microscope. Data collection was performed in step scan mode. For this the sample was first rotated to the starting angle and a predefined number of single shot diffraction images were recorded at this position. Then the sample was rotated by a small increment where again the same predefined number of rotation images was recorded. This procedure was repeated until diffraction images were recorded over the full rotation range. The total accessible rotation range for a given sample is limited by the increasing effective thickness of the sample at higher rotation angles. In our case the overall rotation range was selected such, that for both start and end position still a few diffraction spots were visible on the diffraction images. Data collection parameters of both samples at REGAE are summarized in Tab. S1.

*Muscovite*

Muscovite diffraction data were collected in step scan mode over a rotation range from -60° to +60°. (dataset name: 231215_mica020) The sample was rotated in angle increments of 0.01°. Per rotation increment 12 images were recorded at a repetition rate of REGAE of 12.5 Hz and a bunch charge of 60 fC. The muscovite sample was spanning over the entire aperture of the sample holder with an opening of 0.75 x 3.00 mm, which was defining the accessible lateral size and had a thickness of 670 nm (as independently determined with scanning electron microscope measurements at an edge of the sample).

*Tantalum disulfide (1T-TaS$_2$)*

TaS$_2$ data were collected in step scan mode over a total rotation range from -65° to +65° (dataset name: 240920_TaS2_b09). The sample was rotated in increments of 0.005°. Per rotation increment diffraction patterns from 12 electron pulses were collected and summed up into a single image after application of gain and pedestal correction. REGAE was operating at a frequency of 12.5 Hz with a charge of < 15 fC per shot during data collection. The TaS$_2$ sample had lateral dimensions of 900 x 400 µm$^2$ with a thickness of 30 nm.

Data processing

The JUNGFRAU detector data are stored in HDF5-format. For few frames, either lines or detector segments were apparently not properly written due to overflows in the data back end system. In those cases, the erroneous pixel counts were replaced by values determined by interpolation from neighboring pixels of the same frame and/or adjacent frames using self-written Python scripts. Similarly, dead pixels were replaced by interpolated values. Pedestal and gain mode corrections were applied to the individual diffraction images (*47*). To reduce disk space and simplify downstream data handling all images recorded at every specific rotation angle were summed up into one frame after the corrections. Non-detecting detector segments within the detector outline, i.e. the physical borders between detector modules, were masked and ignored for the data reduction. For the estimation of standard uncertainties, a gain of 10 counts per electron was assumed and the variance was set to 30 counts$^2$ (unit: counts squared) corresponding to the PETS2 (*31*) input parameter '*noiseparameters 10.0 30.0*'.

*Muscovite*

To facilitate data processing of the 12800 muscovite diffraction images, 10 images of neighboring rotation angles were summed up and combined into a single image, which then corresponds to a rotation increment of 0.1°, and exported as unsigned 32-bit TIFF file for subsequent data reduction with PETS2 (*31*). Out of this reduced number of 1280 frames from the muscovite diffraction dataset, 1216 frames were (frame numbers 64 to 1279) used for the data reduction.

During the initial peak search, pattern centres were determined by averaging the coordinates of automatically identified Friedel pairs. A fixed unit cell with parameters $a$ = 5.18 Å, $b$ = 8.99 Å, $c$ = 20.03 Å, $\alpha = \gamma = 90.0°$, and $\beta = 95.76°$ was used to find the initial orientation matrix (*48*). As expected from tests with polycrystalline samples, a strong elliptical distortion was identified and corrected for using the distortion options in PETS2 (*49*). The elliptical amplitude refined to 1.57% with a phase angle of 153.6°. An attempted optimisation of the frame orientation angles did not improve the data reduction and thus the nominal orientation angles were used. Finally, only the pattern centres and the frame-dependent optimisation of the magnification was optimised. The pattern centre moved during the experiment, with the $x$ coordinate being 342.4 (9) pixels, and that of the $y$ coordinate 396.3 (7) pixels. The $x$ coordinate oscillates in a sinusoidal way with an amplitude of about 0.75 pixels and a separation between two maxima of about 120 frames, corresponding to an oscillation frequency of about 0.083 Hz. Integrated intensities were determined using a 2D profile fitting routine assuming a rocking curve width of 0.0 Å$^{-1}$ and a mosaicity of 0.4°. For structure solution and kinematical refinement, frame scales were applied assuming that the Laue class is $2/m$ and standard uncertainties of reflection intensities were determined as recommended in Khouchen *et al*. (*50*). For the dynamical refinement, geometric parameters and integrated reflections intensities were exported without any correction or scale factors. Reflections were bundled into virtual frames so that 28 processed frames with $\Delta\alpha_v = 2.8°$ make up one virtual frame. Subsequent virtual frames overlap by $\Delta\alpha_o = 1.0°$.

*Tantalum disulfide TaS$_2$*

Data treatment was performed analogous to the previous section. Again, 30 frames taken at neighboring rotation angles were summed up for data reduction with PETS2, further reducing the number of frames to be analyzed from 26.000 to 866.(*31*)

The reflections from the peak search could not be indexed with a conventional lattice with 3 basis vectors. The strongest reflections lie on a hexagonal lattice, and all remaining reflections are satellite reflections that can be indexed using two modulation wave vectors $\mathbf{q_1}$ = (0.2450, 0.0681, 1/3) and $\mathbf{q_2}$ = (−0.0681, 0.3131, 2/3). Integrated intensities were determined by simply adding up the pixel intensities within a circular reflection mask. The reflection positions were predicted assuming a rocking curve width of 0.0025 Å$^{-1}$ and a mosaicity of 0.15°.

For structure solution and kinematical refinement, frame scales were applied assuming that the Laue class is $\bar{3}$ and standard uncertainties of reflection intensities were determined as recommended in Khouchen *et al*. (*50*) For the dynamical refinement, geometric parameters and integrated reflections intensities were exported without any corrections. Note that Laue indices consist of five indices *hklmn*. Reflections were bundled into virtual frames so that 20 processed frames with $\Delta\alpha_v = 3.0°$ make up one virtual frame. Subsequent virtual frames overlap by $\Delta\alpha_o = 1.05°$. After the data reduction, in the output files the unit cell parameter $c$, the vector **c*** of the orientation matrix, the modulation wave vectors and the reflection indices *hklmn* were adapted in the output files to achieve a super-centred setting:

$$c' = 3c$$
$$\mathbf{c}^{*\prime} = \frac{1}{3}\mathbf{c}^*$$
$$\mathbf{q}'_1 = (0.2453, 0.0679, 0)$$
$$\mathbf{q}'_2 = (-0.0679, 0.3132, 0)$$
$$l' = 3l + m + 2n$$

In this setting, the super-space structure can be described in a super-centered setting, where the centering vectors are $(0, 0, 0, 0, 0)$, $\left(0, 0, \frac{1}{3}, \frac{2}{3}, \frac{1}{3}\right)$, $\left(0, 0, \frac{2}{3}, \frac{1}{3}, \frac{2}{3}\right)$. Table S2 and Fig. S10 illustrate symmetrical relationships between the 18 modulation wave vectors expressed as linear combinations of $\mathbf{q}'_1$ and $\mathbf{q}'_2$, which are grouped intro 6 groups. Note that in all cases satellite reflections with modulation wave vector $(-m,-n)$ belong to the group $(m,n)$. The absolute length of $|m\mathbf{q}'_1 + n\mathbf{q}'_2|$ is used to order the groups of satellite reflections from lower to higher orders.

## Structure solution and refinement - muscovite

### Structure solution

Based on the output for kinematical refinement, structure solution was successfully attempted with various approaches. Despite the limited completeness of the data set, simple direct methods as implemented in SIR2014 (*33*), charge flipping as implemented in superflip (*34*), and a dual-space approach as implemented in shelxt (*51*) solved the structure (Fig. S11).

### Kinematical refinement

For the subsequent kinematical refinement, it was assumed that the tetrahedral sites are statistically occupied by 25% Al and 75% Si without refining any occupancies. A refinement of site coordinates and isotropic displacement parameters yields $R_{all}$ of 35.1%. This improves significantly by using an extinction correction based on the EXTI-approach (*32*). The final refinement with anisotropic displacement parameters converges with $R_{all}$ of 17.7%. At this stage, the structure model is not charge balanced. In the difference Fourier map, there is one peak at the $2.7\sigma$ level at a distance of 1.01 Å from the O6 site. The free refinement of a hydrogen atom placed at the peak lowers the $R_{all}$ to 17.3% and the O-H distance is 1.07 (3) Å based on the kinematical refinement.

### Dynamical refinement

For the dynamical refinement, the respective output file of PETS2 (*31*) was loaded into Jana2020 (*35*) together with the model from the kinematical refinement without extinction correction. The Bloch wave approach is then used to determine calculated intensities and their derivatives with respect to refinement parameters needed for the least-squares refinement. With the default thickness of 40 nm and optimised scales, $R_{all}$ was 37%. A series of fixed-model refinements tested the dependence of $R_{all}$ on the crystal thickness in the range between 40 nm and 800 nm, revealing a clear minimum at 600 nm (Fig. 3A). With this value, $R_{all}$ dropped to 19%. A subsequent refinement of the model parameters with an increased number of integration steps, $R_{all}$ was further lowered to 14.2%. Finally, the mosaicity of the crystal was modelled as the superposition of incoherently scattering domains representing an isotropic mosaicity of 0.26°. After optimising the orientation angles of the virtual frames, the final refinement with anisotropic displacement parameters converged with $R_{all}$ = 9.9% for 4426 reflections and a refined thickness of 634 (3) nm. Further details on the data set and dynamical refinement are given in Tab. S3.

Apart from the constraints related to Al and Si on tetrahedral sites, no further constraints are used and in the resulting structural model, meaningful anisotropic displacement parameters are obtained (Fig. 3D).

The hydrogen atom present in the structure was initially identified as additional electrostatic potential in the difference Fourier map close to the $3\sigma$ level near the O6 site. In our dynamical structure refinements, we were able to freely refine the position of the hydrogen atom together with anisotropic thermal displacement parameters (ADPs). All refined parameters for the hydrogen atom agree very well with the most recent model from single crystal neutron diffraction measurements within a $2\sigma$ limit (Tab. S4)(*21*). This agreement is also apparent from a visual comparison of the displacement ellipsoids (Fig. 3E).

Structure solution and refinement - Tantalum disulfide 1*T*-TaS$_2$

*Average structure model*

For initial testing purposes, the 3D average structure was determined based on reflections *hkl00*. A Python-script was used to generate an hkl-file with three indices, filtering out satellite reflections. This file was imported with JANA2020. Structure solution in space group $P\bar{3}m$ was straightforward with the three programs used in Section 5.1. A kinematical refinement resulted in very high *R* factors without ($R_{all}$ = 28.5%) and with extinction correction ($R_{all}$ = 24.9%). A dynamical refinement of the unmodulated structure converged with $R_{all}$ = 5.8% ($MR_{all}$ = 4.3%). In the latter model, the anisotropic displacement parameters of sulphur indicated a pronounced displacement parallel to **c**, whereas the displacement ellipsoid of Ta was rather oblate indicating pronounced displacements perpendicular to **c**.

*Structure solution of the modulated structure*

For the determination of the structure in (3+2)d superspace, the transformed output file from the data reduction was imported to JANA2020. Reflections were merged based on point group −3 resulting in 11067 unique reflections. The modulated structure was solved with superflip (*34*). The resulting 5D map is in agreement with the symmetry of the superspace group $X\bar{3}(\alpha,\beta,0)0(-\alpha-\beta,\alpha,0)0$, which is a supercentered setting of the group 147.2.72.1 in the tables by Stokes and Campbell (https://iso.byu.edu/iso/ssg.php) with centering vectors (0,0,0,0,0), (0,0,1/3,2/3,1/3), and (0,0,2/3,1/3,2/3). The solution indicates a strong displacive modulation of the tantalum site within the *ab*-plane, but no displacive modulation along **c**. For the sulphur site, the displacive modulation is pronounced in the *ab*-plane and along **c**.

*Kinematical refinement of the modulated structure*

18 modulation waves (Table S2) for six groups of satellite reflections were defined in JANA2020 (*35*). Based on the structure solution, five parameters for the displacive modulation of the *x* and *y* coordinate of Ta were manually set, accounting for the modulation waves of group 1 and group 2. With this approximate starting model, the star-shaped clustering of Ta atoms in 3D sections of the superspace model is already recognisable. Subsequently, a kinematical refinement was performed using the constraints derived by symmetry, which are set automatically by JANA2020. At first, parameters for the displacive modulation functions of Ta and S for modulation waves of group 1 and group 2 were refined. Then, displacive modulation parameters belonging to the next-higher order group were refined, until the parameters of all defined modulation waves converged. De Wolff-sections of the refined superspace model agree well with sections from the structure solution (Fig. S12). Finally, anisotropic displacement parameters for Ta and S were refined. With extinction correction (EXTI) applied, the final $R_{all}$ is 24.6% for 9924 reflections. For main reflections, $R_{all}$ is 21.6%. A refinement with a resolution limit $\frac{\sin\theta}{\lambda} \leq 0.8$ Å$^{-1}$ improves the results with $R_{all}$ = 19.4% for 4055 reflections.

*Dynamical refinement of the modulated structure*

The final model of the kinematical refinement was used as starting model for the dynamical refinement. For this purpose, the transformed output file from PETS2 was imported with JANA2020. The resolution limit for reflections used in the refinement was set to $\frac{\sin\theta}{\lambda} \leq 0.8$ Å$^{-1}$ to reduce the computational costs for one refinement cycle. Initial parameters for thickness, scales, and correction parameters for the thickness model were taken from the dynamical refinement of the 3D average structure model. The refinement converged quickly, yielding very promising *R* factors for main reflections and stronger satellite reflections. Modulation parameters of anisotropic displacement parameters were refined, but most of them refined to non-significant values so that only the modulation parameters for U33 of the sulphur site were kept in the refinement. Finally, frame orientation angles were optimised before the final round of refinement cycles. Dataset statistics and dynamical structure refinement parameters for the TaS$_2$ sample are provided in Tab. S5. Additionally, refinement *R* factors and selected statistics for the individual satellite reflection groups are provided in Tab. S6.

**Supplementary Text**

Comparative X-ray structure determination of muscovite

*Data collection*

X-ray diffraction data were recorded from exactly the same crystal as used in our MeV ED experiments before. Data were collected at beamline P11 at the PETRA III synchrotron in Hamburg using an X-ray energy of 20 keV. For data collection the sample was continuously rotated at a constant velocity over an angular range from – 60° to + 60°. In total 1200 diffraction images corresponding to an oscillation range of 0.1° per image were recorded on an Eiger 16M detector located 152.9 mm behind the sample.

*Data analysis and structure refinement*

XDS was used for indexing and integration of the X-ray-diffraction data (*52*). Structure refinement using the model from Gatta *et al.* (*21*) was conducted with OLEX2 (*53*). A summary of the X-ray data collection and structure refinement parameter is provided in Tab. S7.

A comparison of the two structures obtained in this work with high-energy MeV ED and X-ray diffraction with neutron data from Gatta *el al.* (*21*) performed with the Bilbao crystallographic server (*54*) is provided in Tab. S8. Whereas the atomic coordinates derived from the ED and X-ray measurements in this work from exactly the same sample agree very well for the non-hydrogen atoms with an average deviation of 0.0094 Å, a slightly larger value of 0.032 Å is obtained for the comparison of our MeV ED data with the neutron reference data. These differences in atomic positions most probably occur from a slightly different composition of the two samples. Whereas we were using synthetically generated muscovite, the sample investigated by Gatta *et al.* (*21*) was naturally occurring muscovite containing mainly iron and other trace metals as impurities giving rise to slightly different atom positions.

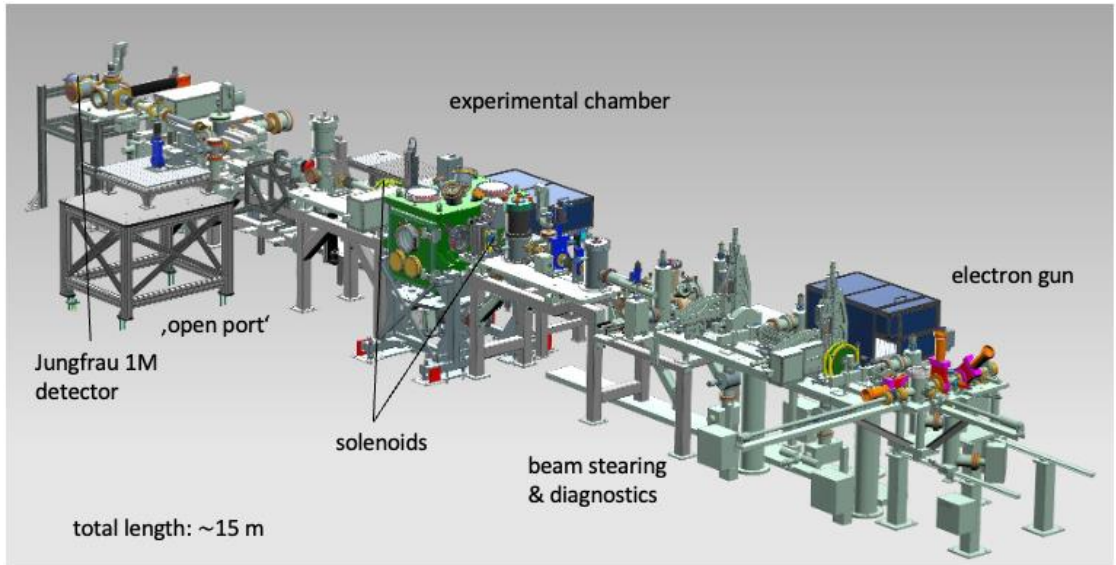

**Fig. S1.**

**Technical overview drawing of the REGAE diffraction facility.** Electrons are emitted from a photo-cathode (right side) and immediately accelerated in a radio-frequency (RF)-cavity to their final energy of 3-5 MeV. A solenoid located shortly before the experimental chamber allows the electron beam to be focused at the interaction point in the experimental chamber, where the sample is located. Diffraction patterns are recorded on a Jungfrau 1M detector (left side) capable of direct electron detection, terminating the experimental setup.

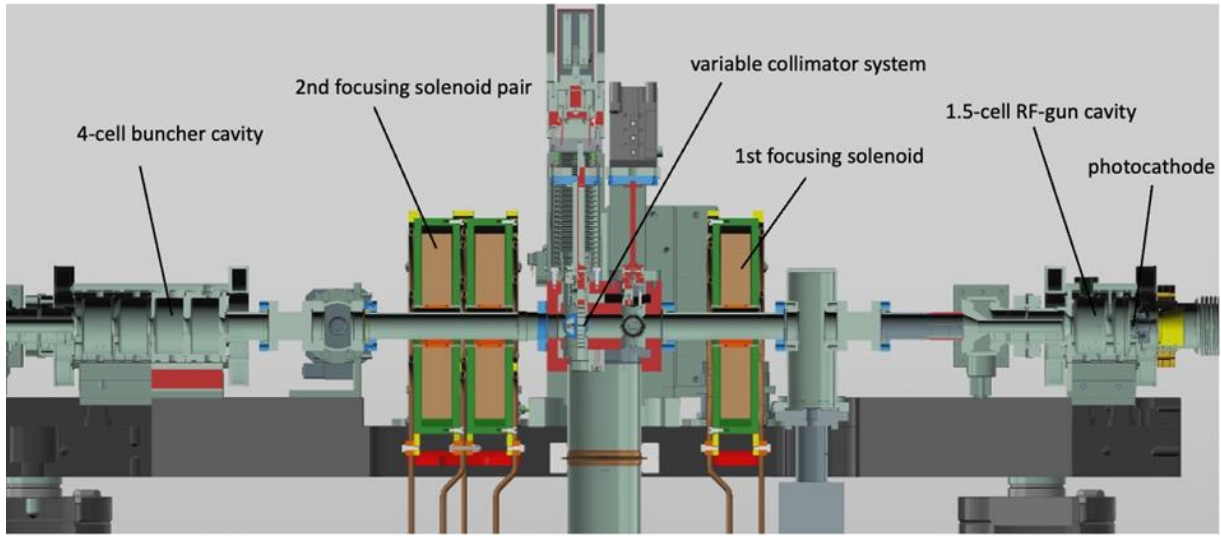

**Fig. S2.**

**Detailed view of the REGAE accelerator.** Technical drawing of the REAGE accelerator part (side view) with labelled key components.

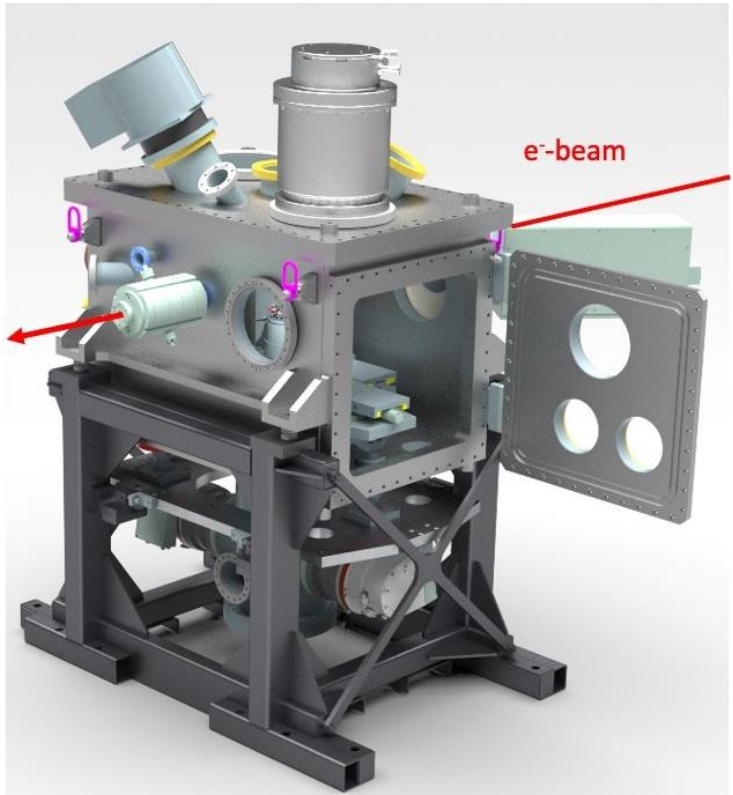

**Fig. S3.**

**Large experimental UHV chamber with inner dimensions of 650 × 590 × 960 mm$^3$ for housing different experiments**. A large front door allows easy exchange of and access to the equipment. The electron beam enters the chamber on the right side of the chamber and exits on the left.

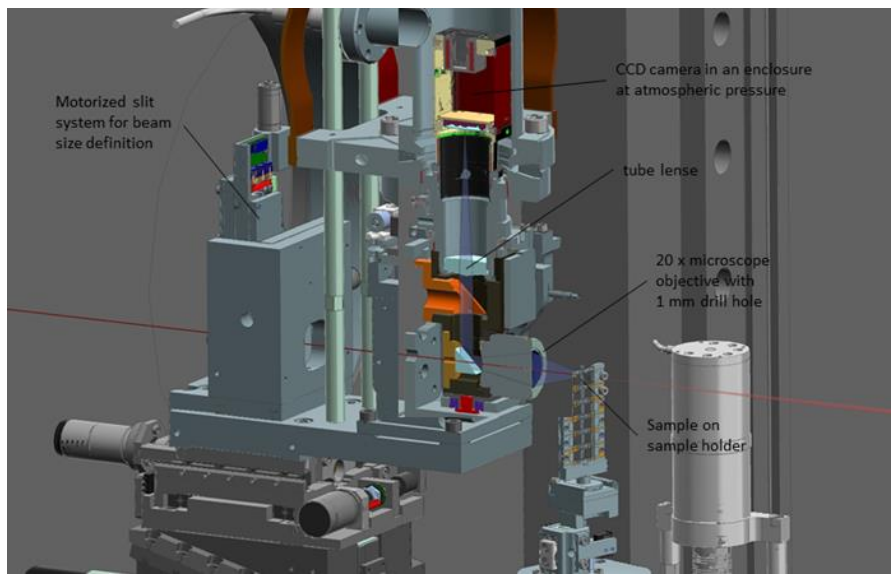

**Fig. S4.**
**Functioning principle of the UHV compatible inline sample viewing microscope at REGAE**. The electron beam passing through the microscope objective is indicated in red, the visible light coming from the sample is deflected upwards and is then recorded on a CCD camera, indicated in blue. An additional optical port allows in-coupling of laser light for pump-probe experiments with the through-the-lens sample excitation either via an optical fiber or by directly guiding the laser light into the optical port.

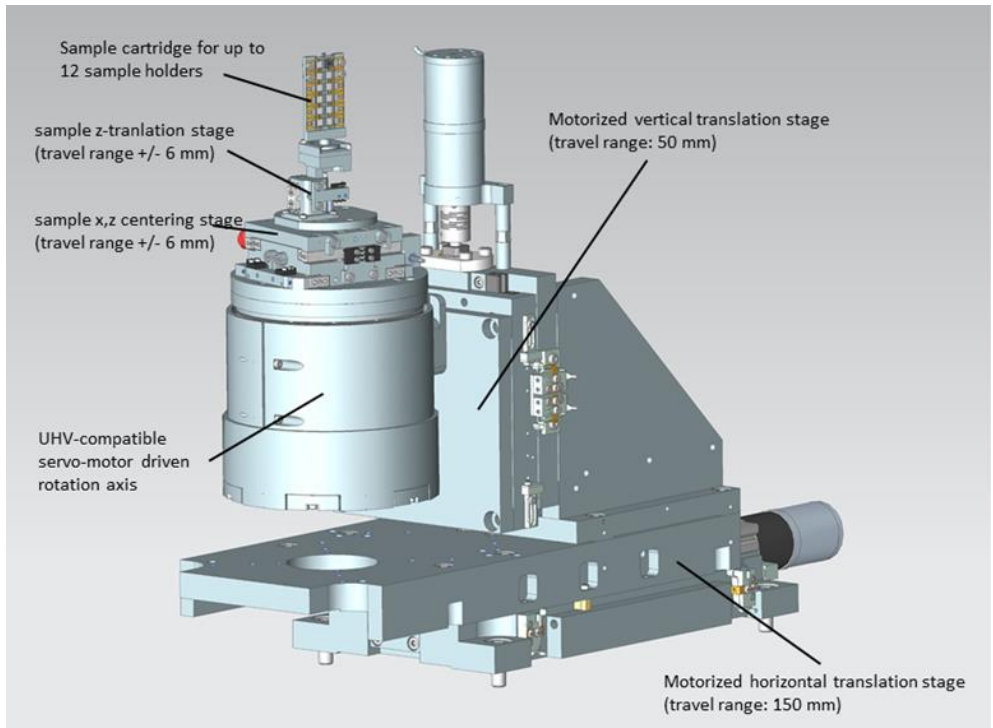

**Fig. S5.**
**Technical drawing of the UHV compatible e--Roadrunner goniometer installed in the REGAE experimental chamber.** The rotation axis is oriented vertically and allows for full 360-degree rotation of the sample with an angular resolution of 0.001°. A centering stage mounted on top of the rotation axis allows precise positioning of the sample in the center of the rotation axis.

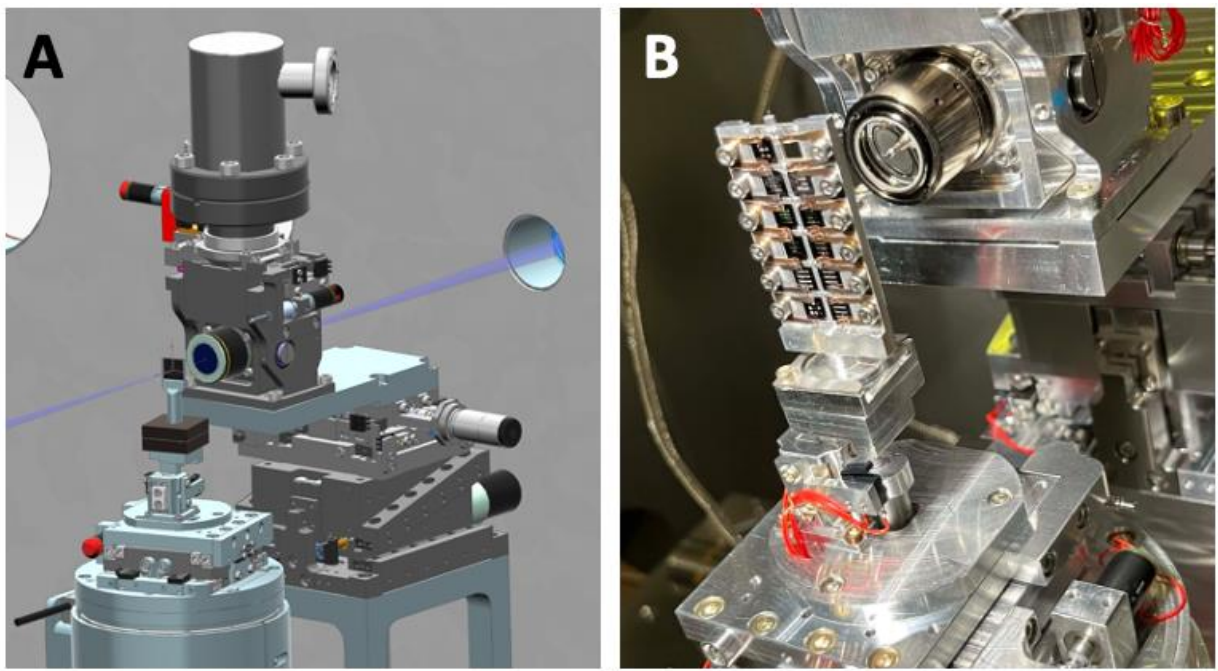

**Fig. S6.**
**Technical drawing and photograph of the UHV-crystallography setup installed at REGAE. (A)** The electron beam (indicated in blue) passes first through the drill hole of the on-axis microscope before it is illuminating the sample located on the crystallography goniometer. (B) Photograph of sample holder at REGAE mounted on the goniometer and loaded with 12 different samples.

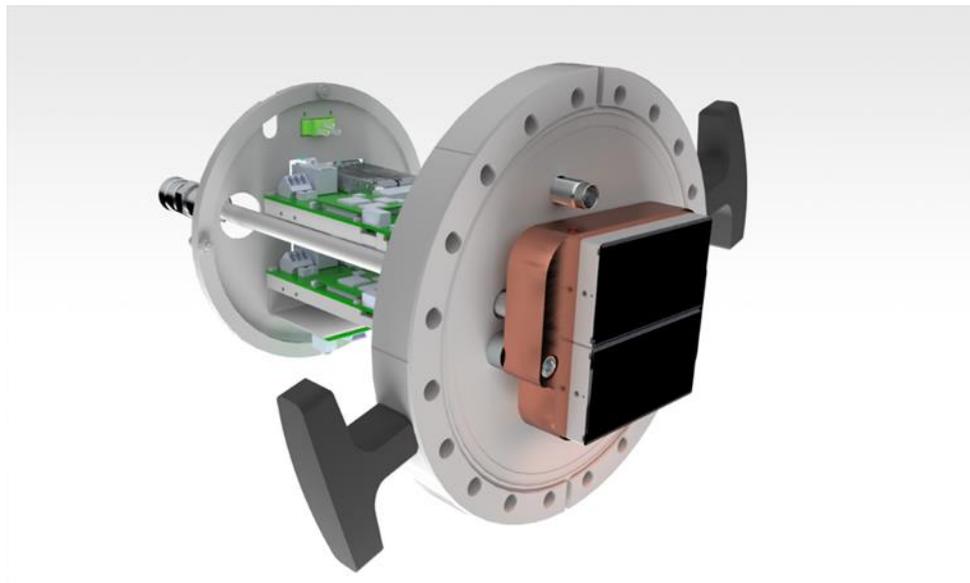

**Fig. S7.**
**Jungfrau detector at REGAE**. UHV compatible Jungfrau detector installed 5.43 m downstream of the sample for direct recording of the MeV single shot diffraction patterns. The detector consists of two modules of each 500 kPixel with a pixel size of 75 µm with a horizontal gap of 3 mm between the two modules.

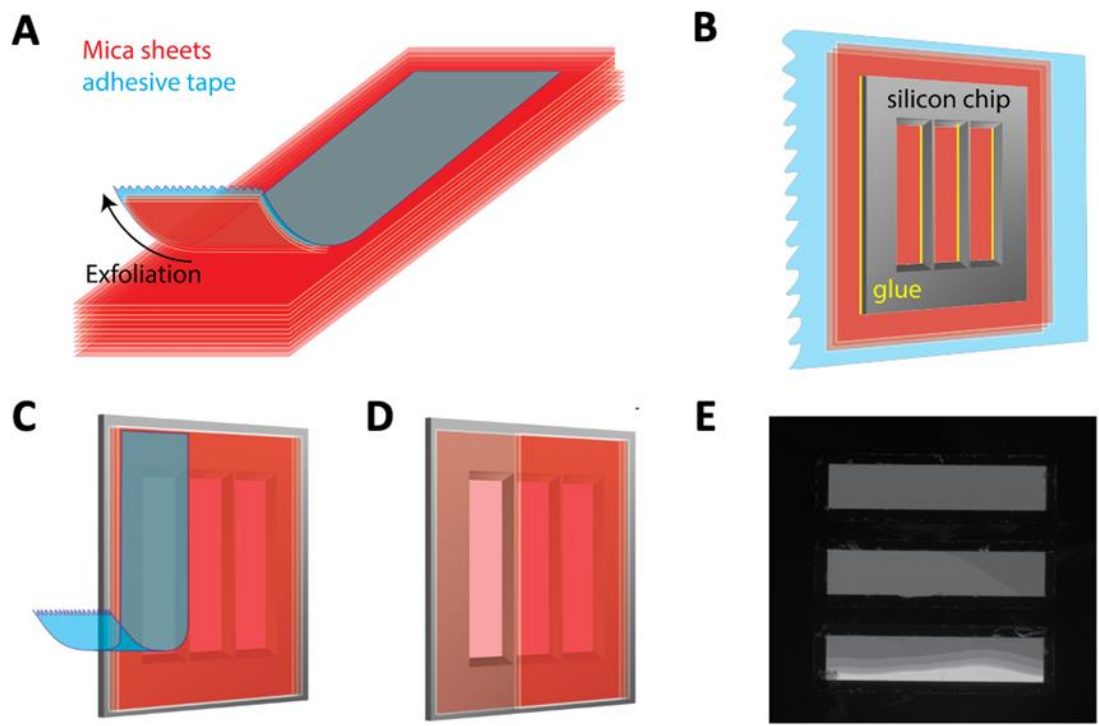

**Fig. S8.**
**Muscovite sample preparation using the technique of exfoliation**. (**A**) a thin sheet consisting of several layers of muscovite is removed from the bulk crystal. (**B**) This thin sheet is then glued onto the REGAE sample holder. (**C**) The sample thickness is further iteratively reduced by removing layers by exfoliation. (**D**) Sample holder with a thinned down sample covering the left of the three rectangular apertures. (**E**) Micrograph of a thinned-down muscovite sample on a REGAE sample frame manufactured from single crystalline silicon.

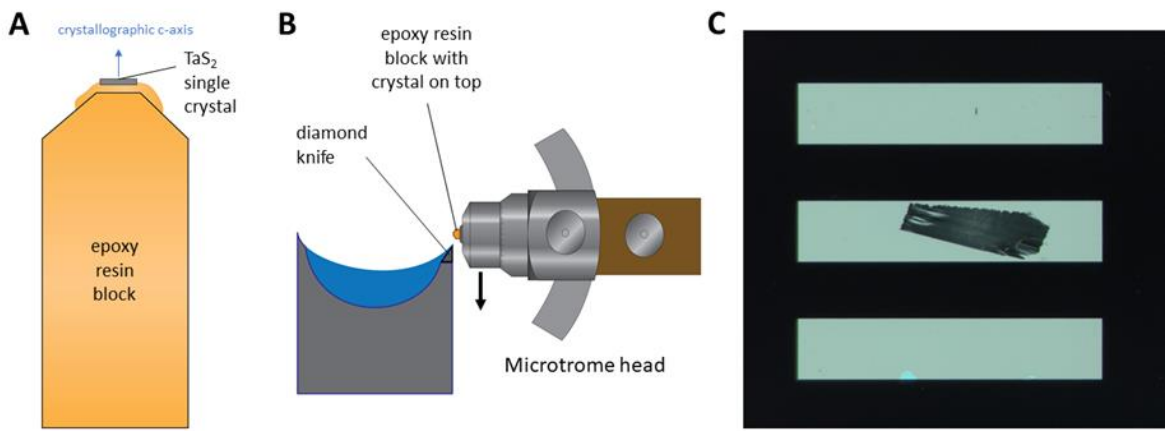

**Fig. S9.**
**1*T*-TaS$_2$ sample preparation using microtome cutting**. (**A**) A 1*T*-TaS$_2$ single crystal is glued on top of an epoxy resin block, so that the c-axis (perpendicular to the 1T-TaS$_2$ layers) is oriented along the long dimension of the resin block. (**B**) The resin block is inserted into the microtome head. For cutting the microtome head moves downwards so that a thin slice (flake) of the sample is cut off or in this case is cleaved off from the supply crystal with a sharp diamond knife. The thin slices / flakes with adjustable thickness are collected in a water bath. (**C**) A 1*T*-TaS$_2$ flake has been 'fished' from the water bath and is mounted on a REGAE sample holder for the diffraction experiment.

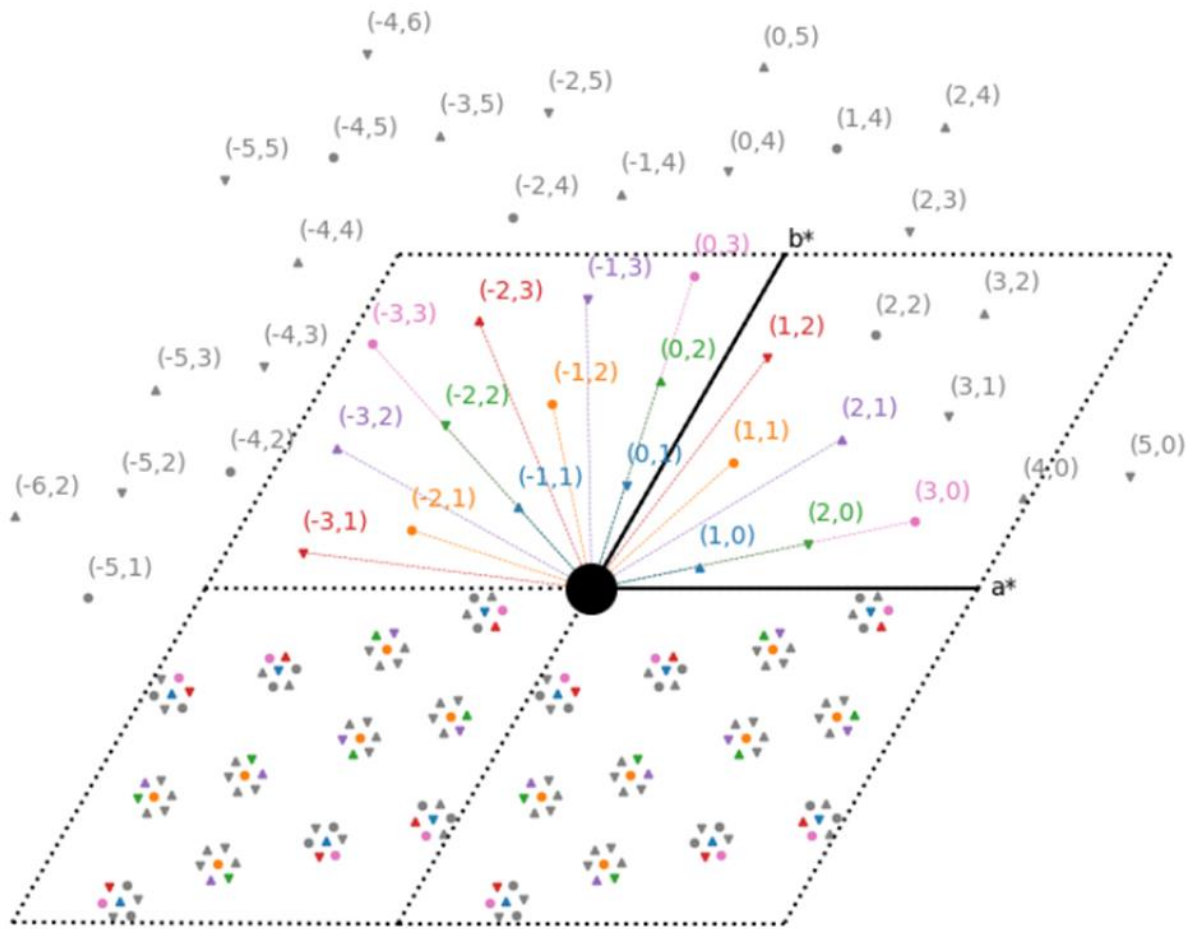

**Fig. S10.**
**Indexing scheme for satellite reflections**. In the upper part the positional relationship between main reflections *hkl00* and satellite reflections *hklmn* is illustrated. In the lower part, satellite reflections from neighboring main reflections are shown. Circles indicate reflections that are present if the condition l = 3n is fulfilled. Triangles pointing upwards are present if the condition l = 3n + 1 is fulfilled. Triangles pointing downwards are present if the condition l = 3n + 2 is fulfilled. Gray satellite reflections are higher-order satellites that are not observed in the data set and not used in the structure analysis. Colored satellite reflections were used for the structure solution and refinement. Colored satellites of the same color belong to the same satellite reflection group. Figure inspired by Fig. 2 in Spijkerman *et al.* (*22*).

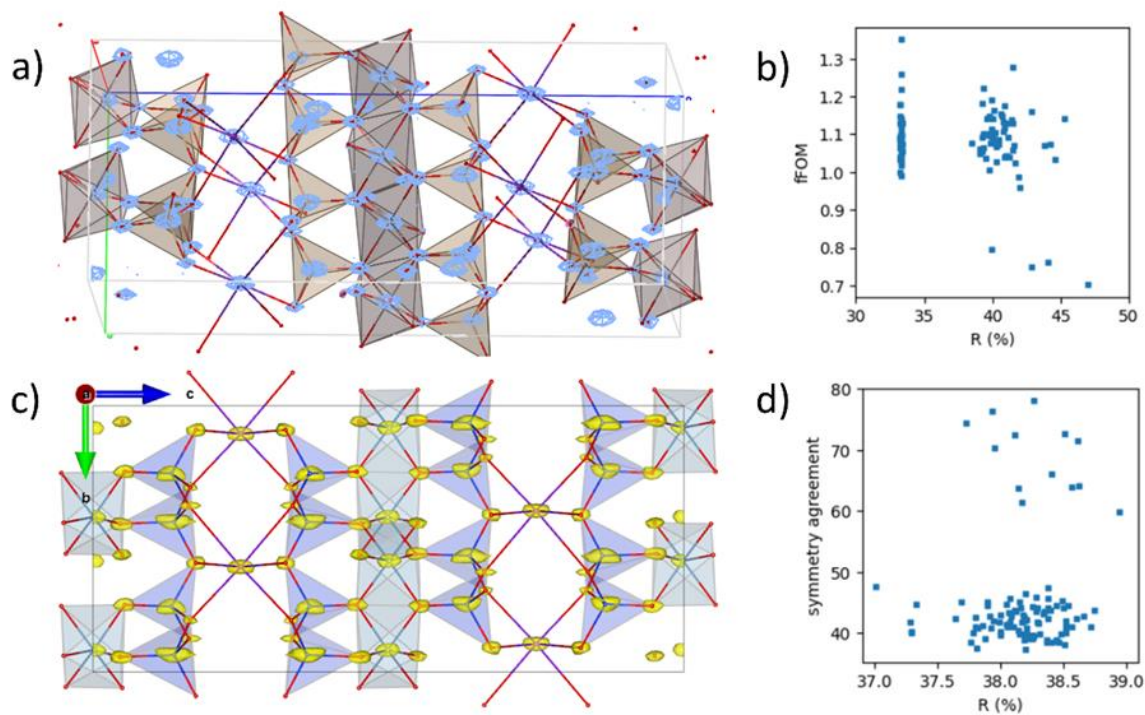

**Fig. S11.**
***ab-initio* structure solution of muscovite from REGAE electron diffraction data**. (**A**) Electrostatic potential map determined with direct methods using SIR2014. A model is overlaid to visualize the structural interpretation of the distribution of peaks. (**B**) Scatter plot, where each data point represents one solution attempt. Out of 100 attempts, 41 solutions have an R factor below 35% and all checked cases correspond to the correct solution. Higher fFOM values are better. (**C**) Electrostatic potential map determined with charge flipping using SUPERFLIP and visualized using VESTA3. A model is overlaid to visualize the structural interpretation of the distribution of peaks. (**D**) Scatter plot, where each data point represents one solution attempt. Lower symmetry agreement factors are better. Almost all solutions with a symmetry agreement factor below 45 provided useful starting models.

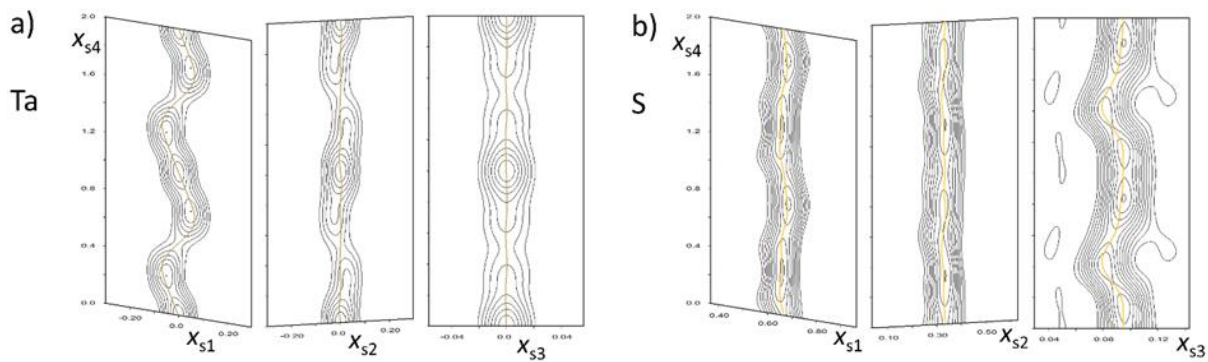

**Fig. S12.**
**The modulated structure of 1*T*-TaS₂**. De Wolff-sections of (**A**) Ta and (**B**) S site with maps from *ab initio* structure solution and overlaid atomic surfaces from kinematical refinement. Contour lines shown in steps of 33 e/Å (**A**) and 3.3 e/Å (**B**) for Ta and S, respectively. The maximum electrostatic potential of Ta is about 260 e/Å, and that of S is about 25 e/Å. $x_{s5}$ was kept at 0 for all sections, so that these sections only represent a small part of the complete structure solution.

**Table S1.**

High-energy electron diffraction data collection parameters of muscovite and 1$T$-TaS$_2$ samples.

| parameter | muscovite | 1$T$-TaS$_2$ |
|---|---|---|
| dataset name | 231215_mica020 | 240920_TaS2_b09 |
| electron energy [MeV] | 3.48(4) | 3.48(4) |
| electron beam diameter [µm] | 250# | 50* |
| pulse charge [fC] | 60 | <15 |
| pulse duration [fs] | ~600 | ~600 |
| rotation range [°] | −60 - +60 | −65 - +65 |
| rotation increment [°] | 0.01 | 0.005 |
| frames per rotation increment | 12 | 12 |
| total number of images | 153.600 | 312.000 |
| number of summed-up diffraction images per rotation increment | 12.800 | 26.000 |
| reduced number of summed up diffraction images for processing with PETS2 (resulting pseudo-rotation range per image) | 1216 (0.1°) | 866 (0.3°) |

# diameter of the full beam (4σ)

*defined by a pinhole 100 mm upstream of the sample

**Table S2.**

Groups of symmetrically related modulation wave vectors.

| Group ID | Group ($m,n$) | symmetrically related wave vectors | | | $\|m\mathbf{q}'_1 + n\mathbf{q}'_2\|$ |
|---|---|---|---|---|---|
| | | #1 | #2 | #3 | |
| 1 | (1,0) | $1\mathbf{q}'_1$ | $1\mathbf{q}'_2$ | $-1\mathbf{q}'_1 + 1\mathbf{q}'_2$ | 0.10 Å$^{-1}$ |
| 2 | (1,1) | $1\mathbf{q}'_1 + 1\mathbf{q}'_2$ | $-1\mathbf{q}'_1 + 2\mathbf{q}'_2$ | $-2\mathbf{q}'_1 + 1\mathbf{q}'_2$ | 0.17 Å$^{-1}$ |
| 3 | (2,0) | $2\mathbf{q}'_1$ | $2\mathbf{q}'_2$ | $-2\mathbf{q}'_1 + 2\mathbf{q}'_2$ | 0.20 Å$^{-1}$ |
| 4 | (1,2) | $1\mathbf{q}'_1 + 2\mathbf{q}'_2$ | $-2\mathbf{q}'_1 + 3\mathbf{q}'_2$ | $-3\mathbf{q}'_1 + 1\mathbf{q}'_2$ | 0.26 Å$^{-1}$ |
| 5 | (2,1) | $2\mathbf{q}'_1 + 1\mathbf{q}'_2$ | $-1\mathbf{q}'_1 + 3\mathbf{q}'_2$ | $-3\mathbf{q}'_1 + 2\mathbf{q}'_2$ | 0.26 Å$^{-1}$ |
| 6 | (3,0) | $3\mathbf{q}'_1$ | $3\mathbf{q}'_2$ | $-3\mathbf{q}'_1 + 3\mathbf{q}'_2$ | 0.29 Å$^{-1}$ |

**Table S3.**

Dataset statistics and dynamical structure refinement parameters for the muscovite sample investigated with the method of high-energy electron diffraction.

| **Crystal / Sample** | |
|---|---|
| Chemical sum formula | $KAl_3Si_3O_{10}(OH)_2$ |
| Z | 4 |
| Crystal system | monoclinic |
| Space group | *C2/c* |
| $a, b, c$ (Å) | 5.20, 9.03, 20.12 |
| $α, β, γ$ (deg.) | 90, 95.8, 90 |
| $V$ (Å³) | 941.4 |
| **Data collection** | |
| Diffractometer | REGAE, DESY, Hamburg |
| Acceleration (kV), $λ$ (Å) | 3480, 0.0031 |
| Detector | Jungfrau 1M |
| Sample temperature (K) | 293 |
| Rotation range $α_{min}, α_{max}, Δα_{step}$ (deg.) | −60.0, +60.0, 0.01 |
| Rotation range per saved frame (deg.) | 0.1 |
| $h_{min}, h_{max}$ | −9, 8 |
| $k_{min}, k_{max}$ | −13, 15 |
| $l_{min}, l_{max}$ | −28, 29 |
| Mosaicity (PETS2) (°) | 0.4 |
| **Data set statistics** | |
| High-resolution limit (Å) | 0.56 |
| Completeness (%) | 66.5 |
| Multiplicity | 2.4 |
| $<I/σ>$ (based on counting statistics) | 16.1 |
| Measured reflections | 4608 |
| Unique reflections | 1968 |
| Unique reflections $I > 3σ$ | 1945 |
| **Dynamical refinement** | |
| virtual frames $Δα_v, Δα_o$ (°) | 2.8, 1.0 |
| No. of scale parameters | 67 (one per virtual frame) |
| No. of structural parameters | 96 |
| No. of refinement parameters | 164 |
| Reflection selection $RSg_{max}, DSg_{min}$ (Å$^{-1}$) | 0.67, 0.005 |
| Bloch waves $g_{max}$ (Å$^{-1}$), $Sg_{max}$ (Å$^{-1}$) | 2.5, 0.01 |
| Refined thickness (nm) | 634 (3) |
| Reflections used in refinement $N_{obs}, N_{all}$ | 4120, 4426 |
| $R_{obs}, R_{all}, wR_{all}$ | 9.8%, 9.9%, 10.8% |
| Post-refinement merged $MN_{obs}, MN_{all}$ | 1756, 1874 |
| $MR_{obs}, MR_{all}, MwR_{all}$ | 9.2%, 9.4%, 10.0% |

**Table S4.**

Comparison of the atomic coordinates and anisotropic thermal displacement parameters for the hydrogen atom of the muscovite structure obtained with MeV ED in our work with reference data from neutron diffraction (ND) experiments by Gatta et al.(*21*).

| Source | T (K) | fractional atomic coordinates | | | | | |
|---|---|---|---|---|---|---|---|
| | | x | | y | | z | |
| ED (this work) | 293 | 0.128(6) | | 0.1508(26) | | 0.4387(28) | |
| ND (*21*) | 295 | 0.1311(7) | | 0.1527(4) | | 0.4420(3) | |
| $\Delta=|p_1-p_2|$ | | $0.5\sigma_\Delta$ | | $0.7\sigma_\Delta$ | | $1.2\sigma_\Delta$ | |
| | | anisotropic thermal displacement parameters (ADPs) | | | | | |
| | | $U_{11}$ (Å$^2$) | $U_{22}$ (Å$^2$) | $U_{33}$ (Å$^2$) | $U_{12}$ (Å$^2$) | $U_{13}$ (Å$^2$) | $U_{23}$ (Å$^2$) |
| ED (this work) | 293 | 0.068(21) | 0.032(12) | 0.17(7) | −0.009(10) | 0.019(28) | 0.052(19) |
| ND (*21*) | 295 | 0.035(2) | 0.034(2) | 0.085(3) | −0.002(1) | 0.018(2) | 0.021(2) |
| $\Delta=|p_1-p_2|$ | | $1.6\sigma_\Delta$ | $0.2\sigma_\Delta$ | $1.2\sigma_\Delta$ | $0.7\sigma_\Delta$ | $0.0\sigma_\Delta$ | $1.6\sigma_\Delta$ |

**Table S5.**

Dataset statistics and dynamical structure refinement parameters for the $TaS_2$ sample investigated with the method of high-energy electron diffraction.

| **Crystal / Sample** | |  |  |
|---|---|---|---|
| Chemical sum formula | $TaS_2$ | | |
| $Z$ | 3 | | |
| Crystal system | Trigonal | | |
| Superspace group | $X\bar{3}(\alpha, \beta, 0)0(-\alpha-\beta, \alpha, 0)0$ | | |
| $a, b, c$ (Å) | 3.3553 (4), 3.3553 (4), 17.6475 (24) | | |
| $\alpha, \beta, \gamma$ (deg.) | 90, 90, 120 | | |
| $V$ (Å³) | 127.06 (4) | | |
| $\mathbf{q}_1, \mathbf{q}_2$ | [0.2453 (2), 0.0679 (3), 0], [−0.0679 (2), 0.3132 (3), 0] | | |
| **Data collection** | | | |
| Diffractometer | REGAE, DESY, Hamburg | | |
| Acceleration (kV), $\lambda$ (Å) | 3480, 0.0031 | | |
| Detector | Jungfrau 1M | | |
| Sample temperature (K) | 293 | | |
| Rotation range $\alpha_{min}, \alpha_{max}, \Delta\alpha_{step}$ (deg.) | −65.0, +65.0, 0.01 | | |
| Rotation range per merged frame (deg.) | 0.15 | | |
| $h_{min}, h_{max}$ | −8, 8 | | |
| $k_{min}, k_{max}$ | −7, 8 | | |
| $l_{min}, l_{max}$ | −31, 29 | | |
| $m_{min}, m_{max}$ | −3, 3 | | |
| $n_{min}, n_{max}$ | −3, 3 | | |
| Mosaicity (PETS2) (°) | 0.15 | | |
| **Data set statistics** | overall | main refl. | satellite refl. |
| High-resolution limit (Å) | 0.45 | | |
| Completeness (%) | 62.4 | | |
| Multiplicity | 3.1 | | |
| $<I/\sigma>$ (based on counting statistics) | 1241.1 | 189.7 | 14.2 |
| Measured reflections | 35016 | 867 | 34091 |
| Unique reflections | 11069 | 296 | 10773 |
| Unique reflections $I > 3\sigma$ | 6391 | 295 | 6096 |
| **Dynamical refinement** | | | |
| virtual frames $\Delta\alpha_v, \Delta\alpha_o$ (°) | 3.0, 1.05 | | |
| No. of virtual frames | 63 | | |
| No. of scale parameters | 63 | | |
| No. of structural parameters | 77 | | |
| No. of refinement parameters | 141 | | |
| Reflection selection $RSg_{max}, DSg_{min}$ (Å⁻¹) | 0.66, 0.0015 | | |
| Bloch waves $g_{max}$ (Å⁻¹), $Sg_{max}$ (Å⁻¹) | 1.8, 0.01 | | |
| Reflections used in refinement $N_{obs}, N_{all}$ | 20629, 11915 | | |
| $R_{obs}, R_{all}, wR_{all}$ | 10.1%, 15.0%, 7.2% | | |
| Post-refinement merged $MN_{obs}, MN_{all}$ | 3430, 4946 | | |
| $MR_{obs}, MR_{all}, MwR_{all}$ | 8.6%, 12.0%, 5.4% | | |

**Table S6.**

Refinement *R* factors and selected statistics grouped by satellite reflection group. Group (0,0) corresponds to the main reflections. $<I/\sigma>$ is the mean value of the intensity to uncertainty ratios of $N_{all}$ reflections.

| group (*m,n*) | $<I/\sigma>$ | $N_{obs}$ | $N_{all}$ | $R_{obs}$ | $R_{all}$ | $wR_{all}$ | $MN_{obs}$ | $MN_{all}$ | $MR_{obs}$ | $MR_{all}$ | $MwR_{all}$ |
|---|---|---|---|---|---|---|---|---|---|---|---|
| (0,0) | 239.4 | 589 | 589 | 4.9% | 4.9% | 5.7% | 136 | 136 | 3.0% | 3.0% | 3.9% |
| (1,0) | 50.1 | 3249 | 3335 | 8.3% | 8.4% | 8.2% | 797 | 804 | 5.9% | 5.9% | 5.2% |
| (1,1) | 16.5 | 2773 | 3400 | 9.6% | 10.6% | 8.9% | 726 | 796 | 7.0% | 7.5% | 5.9% |
| (2,0) | 7.7 | 2113 | 3289 | 11.4% | 14.6% | 11.6% | 636 | 799 | 8.9% | 10.5% | 8.1% |
| (1,2) | 5.6 | 1569 | 3349 | 15.4% | 23.2% | 16.7% | 518 | 807 | 14.0% | 19.4% | 12.6% |
| (2,1) | 2.3 | 898 | 3253 | 28.6% | 40.0% | 33.2% | 343 | 793 | 29.1% | 35.8% | 31.9% |
| (3,0) | 2.0 | 724 | 3414 | 38.2% | 80.8% | 40.4% | 274 | 811 | 47.0% | 80.0% | 40.0% |

**Table S7.**

Dataset statistics and structure refinement parameters for the muscovite X-ray diffraction data recorded at PETRA III beamline P11.

| Crystal / Sample | |
|---|---|
| Chemical sum formula | $KAl_3Si_3O_{10}(OH)_2$ |
| Z | 2 |
| Crystal system | monoclinic |
| Space group | *C*2/*c* |
| Hall group | *-C 2yc* |
| *a*, *b*, *c* (Å) | 5.20, 9.03, 20.10 |
| *α*, *β*, *γ* (deg.) | 90, 95.8, 90 |
| *V* (Å³) | 939.4 |
| **Data collection** | |
| Diffractometer | Beamline P11, PETRA III, DESY, Germany |
| Radiation type | Synchrotron |
| Radiation wavelength (Å) | 0.61991 |
| Detector | Eiger 16M detector, Dectris (Switzerland) |
| Sample temperature (K) | 293 |
| Rotation range $α_{min}$, $α_{max}$ (deg.) | -50, 60 |
| Rotation range per saved frame (deg.) | 0.1 |
| $h_{min}$, $h_{max}$ | -7, 7 |
| $k_{min}$, $k_{max}$ | -12, 11 |
| $l_{min}$, $l_{max}$ | -17, 25 |
| Mosaicity (°) | 21.84 |
| **Data set statistics** | |
| High-resolution limit (Å) | 0.74 |
| Completeness (%) | 89.2 |
| Multiplicity | 2.4 |
| $<I/σ>$ | 41.2 |
| Measured reflections | 2362 |
| Unique reflections | 1039 |
| Unique reflections $I > 3σ$ | 1013 |
| **Refinement parameters** | |
| $R[F^2 > 2σ(F^2)]$, $wR(F^2)$, *S* | 5.0%, 17.9%, 10.6% |
| No. of structural parameters | 105 |
| No. of refinement parameters | 92 |
| H-atom treatment | All H-atom parameters refined |
| $Δρ_{max}$, $Δρ_{min}$ (e Å$^{-3}$) | 1.516, −1.659 |

**Table S8.**

Deviation of atomic coordinates of the muscovite structures measured with MeV electron diffraction (this work), X-ray diffraction (this work), and Neutron diffraction by Gatta *et al.* (*21*).

| Comparison (Bilbao Server) | avg. deviation of atomic positions of non-H (Å) | max. deviation of atomic positions of non-H (Å) | deviation of the atomic position of the H-atom (Å) |
|---|---|---|---|
| ED vs XRD | 0.0094 | 0.024 | 0.1567 |
| ED vs ND | 0.0320 | 0.083 | 0.0619 |
| XRD vs ND | 0.0263 | 0.061 | 0.1754 |